\documentclass[aps,pra,print,twocolumn,showpacs]{revtex4}
\usepackage{amsthm}
\usepackage{amsmath}
\usepackage{amsfonts}
\usepackage{graphicx}
\usepackage{dcolumn}
\usepackage{longtable}
\usepackage{mathrsfs}
\usepackage{setspace}
\usepackage{multirow}
\usepackage{float}
\usepackage{verbatim}
\usepackage{algorithmicx}
\usepackage{algorithm}
\usepackage{algpseudocode}
\usepackage{float}
\usepackage{afterpage}
\newcommand{\ket}[1]{|#1\rangle}
\newcommand{\bra}[1]{\langle#1|}

\newcommand{\Rep}{\operatorname{Re}}
\newcommand{\Imp}{\operatorname{Im}}

\newtheorem{definition}{Definition}

\begin{document}

\title{Optimal arbitrarily accurate composite pulse sequences}

\author{Guang Hao Low, Theodore J. Yoder, and Isaac L. Chuang}
\affiliation{Massachusetts Institute of Technology}
\date{\today}

\begin{abstract}
Implementing a single qubit unitary is often hampered by imperfect control. Systematic amplitude errors $\epsilon$, caused by incorrect duration or strength of a pulse, are an especially common problem. But a sequence of imperfect pulses can provide a better implementation of a desired operation, as compared to a single primitive pulse. We find optimal pulse sequences consisting of $L$ primitive $\pi$ or $2\pi$ rotations that suppress such errors to arbitrary order $\mathcal{O}(\epsilon^{n})$ on arbitrary initial states. Optimality is demonstrated by proving an $L=\mathcal{O}(n)$ lower bound and saturating it with $L=2n$ solutions. Closed-form solutions for arbitrary rotation angles are given for $n=1,2,3,4$. Perturbative solutions for any $n$ are proven for small angles, while arbitrary angle solutions are obtained by analytic continuation up to $n=12$. The derivation proceeds by a novel algebraic and non-recursive approach, in which finding amplitude error correcting sequences can be reduced to solving polynomial equations.
\end{abstract}


\maketitle

\section{Introduction}

Quantum computers are poised to solve a class of technologically relevant problems intractable on classical machines \cite{Nielsen2004}, but scalable implementations managing a useful number of qubits are directly impeded by two general classes of errors \cite{Merrill2012}. On one hand, unwanted system-bath interactions in open quantum systems lead to decoherence and, on the other, imperfect controls for addressing and manipulating qubit states result in cumulative errors that eventually render large computations useless.

Systematic amplitude errors, the consistent over- or under- rotation of a single-qubit unitary operation by a small factor $\epsilon$, are one common control fault. The discovery of a protocol for the complete and efficient suppression of these errors would greatly advance the field of quantum control with applications as far ranging as implementing fault-tolerant quantum computation and improving nuclear magnetic resonance spectra acquisition. Due to the broad scope of systematic amplitude errors, this problem has been attacked repeatedly by a variety of methods with varying degrees of success \cite{Tycko1985,Levitt1986,Brown2004,Ichikawa2012,Jones2013,Husain2013}. A concept common to most approaches is the composite pulse sequence, in which some number of $L$ carefully chosen erroneous primitive unitary operations, or pulses, are applied successively such that a target ideal rotation is approximated to some order $n$ with an exponentially reduced error $\mathcal{O}(\epsilon^{n+1})$.

In the realm of quantum computation, the criteria for useful pulse sequences are stringent:  (1) For each order $n$, a procedure for constructing a pulse sequence correcting to that order is known. (2) This construction gives sequence lengths $L$ that scale efficiently with $n$, that is $L=\mathcal{O}(n^k)$ with $k$ as small as possible. (3) Sequences should be `fully-compensating' or `Class A' \cite{Levitt1986}, meaning they operate successfully on arbitrary and unknown states (in contrast to `Class B' sequences that only operate successfully on select initial states). (4) Although finite sets of universal quantum gates exist \cite{Nielsen2004}, ideally sequences should be capable of implementing arbitrary rotations so that quantum algorithms can be simplified conceptually and practically.

One finds that there are currently no sequences satisfying all four of these criteria and suppressing systematic amplitude errors. In the literature, SCROFULOUS \cite{Cummins2003}, PB$_1$, BB$_1$ \cite{Wimperis1994} satisfy criteria (3) and (4) but offer corrections only up to order $n=2$. Unfortunately, generalizations of these to arbitrary $n$ and come with prohibitively long sequence lengths, so that criterion (2) ends up unsatisfied. Typically, a sequence correct to order $n+1$ is recursively constructed from those at order $n$, resulting in an inefficient sequence length $L=2^{\mathcal{O}(n)}$ \cite{Brown2004}, although numerical studies suggest that efficient sequences with $L=\mathcal{O}(n^{3.09})$ exist \cite{Brown2004}. To date, other classes of systematic control errors \cite{Viola1999, Khodjasteh2010,Souza2012} do not fare better.

There are some provable successes in efficient pulse sequences, though. However, to find them, one must relax criterion (4) that requires arbitrary rotations. For example, if one restricts attention to correcting $\pi$ rotations in the presence of amplitude errors, Jones proved the impressive result that sequences with $L=\mathcal{O}(n^{1.47})$ \cite{Jones2013,Tycko1985} are possible. Uhrig efficiently implements the identity operator in the presence of dephasing errors with $L=\mathcal{O}(n)$ \cite{Uhrig2007}. If we also relax the  criterion (3) and settle for specialized Class B sequences that take $\ket{0}$ to $\ket{1}$ (those we call inverting sequences), Vitanov has found efficient narrowband sequences for amplitude errors also with $L=\mathcal{O}(n)$ \cite{Vitanov2011}. Notably, both Uhrig's and Vitanov's results were achieved via algebraic, non-recursive processes. In fact, as we show, a more generalized algebraic approach in the amplitude error case can reinstate the crucial criteria (3) and (4), while maintaining Vitanov's efficient length scaling.

Our main result is exactly such an algebraic generalization, a non-recursive formalism for systematic amplitude errors. With this, we prove a lower bound of $L=\mathcal{O}(n)$ for Class A sequences comprised of either primitive $\pi$ or $2\pi$ rotations, then constructively saturate this bound to a constant factor with $L=2n$ (plus a single initializing rotation). The improvement of these new sequences over prior state-of-the-art is illustrated in Table~\ref{sequence_table}. We derive optimal closed-form solutions up to $n=4$ for arbitrary target angles, and perturbative solutions for any $n$, valid for small target angles. We then analytically continue these perturbative solutions arbitrary angles up to $n=12$. Since any random or uncorrected systematic errors in the primitive pulses accumulate linearly with sequence length, optimally short sequences such as ours minimize the effect of such errors.

We define the problem statement for amplitude-error correcting pulse sequences mathematically in Section \ref{Pulse sequences}, leading, in Section \ref{section_Pulse sequence constraint equations}, to a set of constraint equations that such pulse sequences must satisfy, which is then solved in Section \ref{section_Solving the constraints} by three approaches: analytical, perturbative, and numerical. The analytical method is interesting as it gives closed form solutions for low order sequences in a systematic fashion. The perturbative method relies on invertibility of the Jacobian of the constraints and is used for proving the existence of solutions for select target angles. The numerical method is the most straightforward and practical for higher orders, giving optimally short pulse sequences for correction orders up to $n=12$. Section \ref{section_Further extensions} then presents several generalizations of our results, including discussions on narrowband toggling, nonlinear amplitude errors, random errors, and simultaneous correction of off-resonance errors. Finally, we point out differences and similarities between our sequences and existing art in Section \ref{section_Comparison with prior art}, and conclude in Section \ref{section_Conclusion}.

\section{Pulse sequences}
\label{Pulse sequences}

A single qubit rotation of target angle $\theta_T$ about the axis $\vec n$ is the unitary $R_{\vec n}[\theta_T]=\exp\left(-i\theta_T(\vec n\cdot\vec\sigma)/2\right)$, where $\vec\sigma=(\hat X,\hat Y,\hat Z)$ is the vector of Pauli operators. Without affecting the asymptotic efficiency of our sequences, Euler angles allow us to choose $n_z=0$, and consequently we define $R_{\varphi}[\theta_T]=\exp\left(-i\theta_T\hat\sigma_{\varphi}/2\right)$ for ${\hat\sigma_{\varphi}=\hat X\cos\varphi+\hat Y\sin\varphi}$. However, we only have access to imperfect rotations ${M_{\varphi}[\theta]=R_{\varphi}[(1+\epsilon)\theta]}$ that overshoot a desired angle $\theta$ by $\epsilon\theta$, $|\epsilon|\ll1$. With these primitive elements, we construct a pulse sequence $\mathcal{S}$ consisting of $L$ faulty pulses:
\begin{equation}
\mathcal{S}=M_{\varphi_1}[\theta_1]M_{\varphi_2}[\theta_2]\dots M_{\varphi_L}[\theta_L].
\end{equation}
Denote by $\vec\varphi$ the vector of phase angles $(\varphi_1,\varphi_2,\dots,\varphi_L)$, which are our free parameters. Leaving each amplitude $\theta_j$ as a free parameter ({\it e.g.} SCROFULOUS \cite{Cummins2003}) may help reduce sequence length, but we find that a fixed value $\theta_j=\theta_0$ leads to the most compelling results.

The goal is to implement a target rotation $R_{\varphi_0}\left[\theta_T\right]$ (or, without loss of generality, $R_0\left[\theta_T\right]$ by the replacement $\varphi_j\rightarrow\varphi_j-\varphi_0$) including the correct global phase, with a small error. The trace distance \cite{Nielsen2004}
\begin{equation}
D(\hat{U},\hat{V}) = \|\hat{U}-\hat{V}\| = \frac{1}{2}\text{Tr}\sqrt{\left(\hat{U}-\hat{V}\right)^{\dag}\left(\hat{U}-\hat{V}\right)}
\end{equation}
is a natural metric for defining errors between two operators $\hat{U}$, $\hat{V}$ \cite{Brown2004}. We demand that the pulse sequence implements
\begin{equation}\label{bb_constraint1}
\mathcal{S}=R_0\left[-\epsilon\theta_T\right]+\mathcal{O}(\epsilon^{n+1}),
\end{equation}
so that the corrected rotation $U_T=\mathcal{S}\cdot M_0\left[\theta_T\right]=R_0\left[\theta_T\right]+\mathcal{O}(\epsilon^{n+1})$ has trace distance with the same small leading error $D(U_T,R_0\left[\theta_T\right]) = \mathcal{O}(\epsilon^{n+1})$. Thus constructed, $U_T$ implements $R_0\left[\theta_T\right]$ over a very wide range of $\epsilon$ due to its first $n$ derivatives vanishing and so has broadband characteristics \cite{Merrill2012}. 

For completeness, we mention other error quantifiers. First, is the fidelity $F(\hat{U},\hat{V})=\|\hat{U}\hat{V}^{\dag}\|$, which is not truly a distance metric, but can be easier to compute and bounds $1-F(\hat{U},\hat{V})\le D(\hat{U},\hat{V}) \le \sqrt{1-F(\hat{U},\hat{V})^2}$ \cite{Nielsen2004}. The infidelity of $U_T$ is then $1-F(U_T,R_0\left[\theta_T\right])=\mathcal{O}(\epsilon^{2n+2})$, which is a commonly used quantifier \cite{Merrill2012,Jones2013}. Finally, for the specialized Class B sequences called inverting sequences the transition probability $|\bra{1}\hat U\ket{0}|^2$ is a viable quantity for comparison \cite{Vitanov2011}.

\section{Constraint equations}
\label{section_Pulse sequence constraint equations}
We now proceed to derive a set of equations, or constraints, on the phase angles $\vec\varphi$ that will yield broadband correction. We begin very generally in the first subsection by assuming just $\theta_j=\theta_0$ as mentioned before, but then we specialize in the subsequent two subsections to the case $\theta_0=2\pi$ and the case of symmetric sequences, both of which greatly enhance tractability of the problem.

\subsection{Equal amplitude base pulses}

To begin, we obtain an algebraic expression for $\mathcal{S}$ by a direct expansion of a length $L$ sequence. Defining $\theta_0'=(1+\epsilon)\theta_0/2$,
\begin{align}
\mathcal{S}&=\prod_{j=1}^L M_{\varphi_j}[\theta_0]=
\cos^{L}\left(\theta_0'\right)\prod_{j=1}^L\left( 1-i \tan\left(\theta_0'\right)\hat{\sigma}_{\varphi_j}\right)\\\nonumber
&=\sum_{j=0}^L A_L^j\left(\theta_0'\right)\hat\Phi_L^j(\vec\varphi),
\end{align}
where indices in the matrix product ascend from left to right, $A_L^j(s)=(-i)^j\sin^j\left(s\right)\cos^{L-j}\left(s\right)$, and $\hat\Phi_L^j$ are noncommutative elementary symmetric functions generated by $\prod_{j=1}^L \left(1+t\hat\sigma_{\varphi_j}\right)=\sum_{j=0}^L t^j\hat\Phi_L^j$ \cite{Gelfand1995}. The $\hat\Phi_L^j$ are hard to work with so by applying the Pauli matrix identity
\begin{align}
\sigma_{\varphi_1}\sigma_{\varphi_2}\dots\sigma_{\varphi_j}=\exp\left(i\hat Z\sum_{k=1}^j(-1)^k\varphi_k\right)\hat X^j,
\end{align}
we obtain a more useful expression as functions of the phase angles $\varphi_j$:
\begin{align}
\hat\Phi_L^j(\vec\varphi)&=\left(\Rep[\Phi_L^j(\vec\varphi)]I-i\Imp[\Phi_L^j(\vec\varphi)]\hat Z\right)\hat X^j,\\\label{phase_sum}
\Phi_L^j(\vec\varphi)&=\sum_{1\le h_1<h_2<\dots<h_j\le L}\exp\left(-i\sum_{k=1}^j(-1)^k\varphi_{h_k}\right).
\end{align}
By defining the terminal case $\Phi_L^0(\vec\varphi)\equiv 1$, the phase sums $\Phi_L^j$ are efficiently computable at numeric values of the phases by the recursion $\Phi_L^j=\Phi_{L-1}^j+\Phi_{L-1}^{j-1}e^{i(-1)^{j+1}\varphi_L}$ using dynamic programming ({\it i.e.} start from the terminal case and fill in the table $\Phi_L^j$ for all desired $j$ and $L$).

\begin{table}
  \begin{tabular}{ l c l }
    Name & Length & Notes \\ \hline
    \hline
    SCROFULOUS & $3$ & $n=1$, non-uniform $\theta_j$ \cite{Cummins2003} \\ \hline
    P$n$, B$n$ & $\mathcal{O}(e^{n^2})$ & Closed-form \cite{Brown2004} \\ \hline
    \multirow{2}{*}{SK$n$} & \multirow{2}{*}{$\mathcal{O}(n^3)$}
              & $n\le30$,    numerical \cite{Brown2004}   \\
              & & $n>30$, conjectured     \\ \hline
    \multirow{3}{*}{AP$n$ (PD$n$)} & \multirow{3}{*}{$2n$}
           & $n\le 3(4)$, closed-form  \\
           & & $n\le 12$, analytic continuation \\
           & & $n>12$, conjectured \\ \hline
    \multirow{1}{*}{ToP$n$}& \multirow{1}{*}{$2n$}
           & arbitrary $n$, perturbative \\ \hline
  \end{tabular}
  \caption{Comparison of known pulse sequences operating on arbitrary initial states that suppress systematic amplitude errors to order $n$ for arbitrary target rotation angles. Arbitrary accuracy generalizations are known or conjectured for all with the exception of SCROFULOUS. The sequences AP$n$, PD$n$, and ToP$n$ are presented in this work. Of interest is the subset of AP$n$ sequences labeled ToP$n$ for which arbitrary accuracy is provable perturbatively for small target angles.}
  \label{sequence_table}
\end{table}

Combining the expansion of $\mathcal{S}$ with Eq.~\eqref{bb_constraint1} then imposes a set $\mathcal{B}_{n,L}$ of real constraints on $\vec\varphi$ to be satisfied by any order $n$, length $L$ sequence. $\mathcal{B}_{n,L}$ is obtained by first matching coefficients of the trace orthogonal Pauli operators on either side of Eq.~\eqref{bb_constraint1}. We then obtain in terms of normalized error $x=\frac{1}{2}\epsilon \theta_0$ and normalized target angle $\gamma=\theta_T/\theta_0$ the necessary and sufficient conditions
\begin{equation}\label{bb_constraint2a}
\sum_{j\in
\big\{
\begin{smallmatrix}
\text{even}\\
\text{odd}
\end{smallmatrix}
}A_L^j\left(\theta_0/2+x\right)\Phi_L^j(\vec\varphi)
=\big\{
\begin{smallmatrix}
\cos(x\gamma)\\
i\sin(x\gamma)
\end{smallmatrix} +\mathcal{O}(x^{n+1}).
\end{equation}
Second, the complex coefficients of $x^0, x^1,\dots,x^n$ are matched, giving $2(n+1)$ complex equations linear in the phase sums $\Phi_L^j(\vec\varphi)$, or $|\mathcal{B}_{n,L}|=4(n+1)$ real constraints.

However, these constraints $\mathcal{B}_{n,L}$ are intractable to direct solution, and a simplifying assumption is necessary. It should be reasonable to suspect that the small rotation $R_0\left[-\epsilon\theta_T\right]$ can be generated by small pure error terms $R_{\varphi}\left[\epsilon \theta_{0}\right]$. We will therefore set $\theta_0=2\pi$ \cite{Brown2004}. Note that $\theta_0=\pi$ is also a tractable case but is related to the $2\pi$-pulse case by phase toggling and so need not be considered separately. We give more detail on toggling in Section \ref{section_Further extensions}.

\subsection{Assuming base pulses of $\theta_0=2\pi$}
We now enumerate several key results, due simply to imposing $\theta_0=2\pi$, that apply to all order $n$, length $L$, $2\pi$-pulse sequences. First, Eq.~\eqref{bb_constraint2a} reduces to
\begin{equation}\label{bb_constraint_2pi}
(-1)^L\sum_{j=0}^L A_L^j(x)\Phi_L^j(\vec\varphi)=e^{ix\gamma}+\mathcal{O}(x^{n+1}),
\end{equation}
by summing its even and odd parts, justified by noting ${A_L^j(\theta_0/2+x)\rightarrow (-1)^L A_L^j(x)}$ hence $x$ occurs only in even (odd) powers for $j$ even (odd). By matching coefficients of powers of $x$, this represents $2(n+1)$ real constraints. Second, the $x^0$ terms in Eq.~\eqref{bb_constraint_2pi} match if and only if $L$ is even. Assuming this, $2n$ constraints remain. Third, we arrive at our most important result by transforming Eq.~\eqref{bb_constraint_2pi} with the substitution $x\rightarrow i\tanh^{-1}(y)$. This eliminates trigonometric and exponential functions from Eq.~\eqref{bb_constraint_2pi}, and (assuming $L$ is even) leaves
\begin{align}
\left[(1-y)(1+y)\right]^{-L/2}\sum_{j=1}^Ly^j\Phi_L^j(\vec\varphi)=\left[\frac{1-y}{1+y}\right]^{\gamma/2}.
\end{align}
Upon rearrangement, this is a generating equation for values that the phase sums $\Phi_L^j(\vec\varphi)$ must satisfy. The functions $f^j_L(\gamma)$ generated by  $\sum_{j=0}^\infty f_L^j(\gamma)y^j=(1+y)^{(L-\gamma)/2}(1-y)^{(L+\gamma)/2}$ are, in fact, real polynomials in $\gamma$ of degree $j$ which generalize those of Mittag-Leffler \cite{Bateman1940}. We can now write
\begin{align}\label{phase_sum=ffunction}
\Phi_L^j(\vec\varphi)&=f^j_L(\gamma),\quad 0< j\le n,\\\label{ffunction}
f^j_L(\gamma)&=\sum_{k=0}^j(-1)^k\binom{T}{k}\binom{L-T}{j-k},\,T\equiv\frac12(\gamma+L)
\end{align}
Eq.~\eqref{phase_sum=ffunction} is, in our opinion, the simplest and most useful representation of the non-linear (in $\vec{\varphi}$) constraints that form the basis for our solutions.

In our notation the leading error of an order $n$, even $L$, $2\pi$-pulse sequence $\mathcal{S}_{2\pi}$ has a simple form,
\begin{equation}\label{leading_error}
\mathcal{S}_{2\pi}\cdot R_{0}[2x\gamma]=I-\left(f_L^{n+1}(\gamma)\hat X^{n+1}-\hat\Phi_L^{n+1}(\vec\varphi)\right)(-ix)^{n+1}.
\end{equation}
Now, we recognize the operator on the right of Eq.~\eqref{leading_error} must be unitary. Thus, if a set $\vec\varphi$ satisfies Eq.~\eqref{phase_sum=ffunction} for ${0< j < k}$ for any even integer $k$, $\Rep[\Phi_L^k(\vec\varphi)]=f_L^k(\gamma)$ follows automatically. So we define $\mathcal{B}_{n,L}^{2\pi}$, the set of constraints resulting from applying $\theta_0=2\pi$ to $\mathcal{B}_{n,L}$, to consist of the $n$ complex equations from Eq.~\eqref{phase_sum=ffunction} ignoring the real parts for even $j$.
\begin{equation}\label{constraints}
\mathcal{B}_{n,L}^{2\pi}=\left\{\begin{array}{ll}\vspace{3pt}
\Rep\Phi^j_L(\vec\varphi)=f_L^j(\gamma)&\text{,  }j\text{ odd}\\
\Imp\Phi^j_L(\vec\varphi)=0&\text{,  for all }j\\
\end{array}\right\}_{j=1,2,\dots,n}
\end{equation}
Thus, $|\mathcal{B}_{n,L}^{2\pi}|=\lceil3n/2\rceil$.

In fact, it is not difficult to place a lower bound on the pulse length $L$ for a sequence correcting to order $n$ using the framework we have so far. This is the first bound of its kind, and, given our solutions of the constraints to come in section \ref{section_Solving the constraints}, it must be tight to a constant factor. Begin the argument by way of contradiction, letting $n>L$. In examining $\mathcal{B}_{n,L}^{2\pi}$, observe $\Phi_L^j(\vec\varphi)=0$ for ${L<j\le n}$, but $f^j_L(\gamma)$ is a real polynomial in $\gamma$ of degree $j$. Hence $0=\Phi^n_L(\vec\varphi)=f^n_L(\gamma)$ cannot be satisfied for arbitrary $\gamma$. Likewise, if $L=n$, then $1=|\Phi_L^n(\vec\varphi)|=|f_L^n(\gamma)|$ cannot be satisfied for arbitrary $\gamma$. Thus $L>n$ is necessary.

\subsection{Assuming phase angle symmetries}

Some constraints in $\mathcal{B}_{n,L}^{2\pi}$ can be automatically satisfied if appropriate symmetries on the phase angle are imposed. A symmetry property of the phase sums, $\Phi_L^j(\vec\varphi)=[\Phi_L^j((-1)^j\vec\varphi_R)]^{*}$ with reversed phase angles $\vec\varphi_R=(\varphi_L,\varphi_{L-1},...,\varphi_1)$, motivates us to impose a palindromic (antipalindromic) symmetry on the phases, $\vec\varphi=+\vec\varphi_{R}$ ($\vec\varphi=-\vec\varphi_R$) so that $\Imp[\Phi_L^j(\vec\varphi)]=0$ for even (odd) $j$. Removing these equations from $\mathcal{B}_{n,L}^{2\pi}$, we are left with the subset $\mathcal{B}_{n,L}^{\text{PD}}$ ($\mathcal{B}_{n,L}^{\text{AP}}$).
By definition, $\varphi^{\text{AP}}_k=-\varphi^{\text{AP}}_{L-k+1}$ and $\varphi^{\text{PD}}_k=\varphi^{\text{PD}}_{L-k+1}$. In both cases, we have $|\mathcal{B}_{n,L}^{\text{PD}}|=|\mathcal{B}_{n,L}^{\text{AP}}|=n$ real constraints to be satisfied by $\lceil L/2\rceil$ real variables $\vec\varphi^{\text{AP}}$ or $\vec\varphi^{\text{PD}}$. With what minimum $L$ is this possible, and is it of the linear length scaling $L=\mathcal{O}(n)$ suggested by our lower bound?

\section{Solving the constraints}
\label{section_Solving the constraints}

\begin{figure}
\includegraphics[width=\columnwidth]{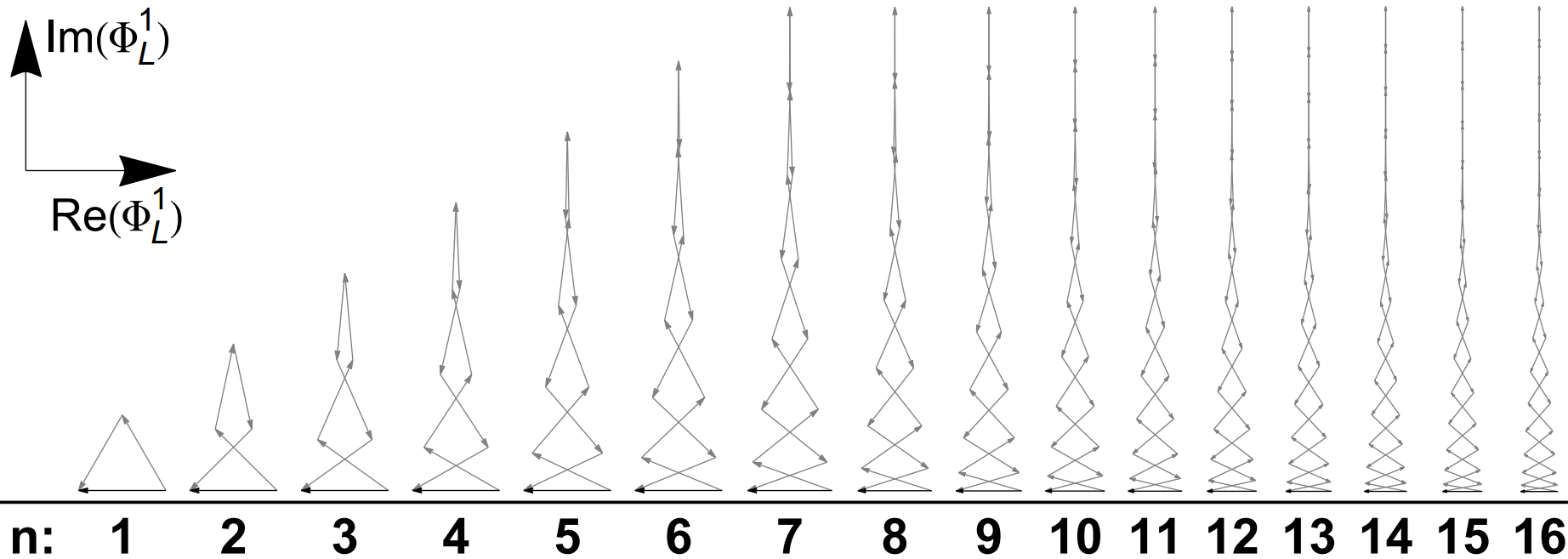}
\caption{\label{APLiePlot}
Pulse sequences with $\theta_0=2\pi$ can be visualized with phasor diagrams as the first constraint equation, $\Phi^1_L(\vec\varphi)=\sum_{k=1}^{L}e^{i\varphi_k}=f^1_L(\gamma)=-\gamma$ is a sum of phases. As examples, the phase angles $\left.\vec{\varphi}_L\right|_{\gamma=1}=(\varphi_1,...,\varphi_L)$ are plotted tip-to-tail to scale for ToP$n$ sequences where $L=2n$ and all arrows are of unit length. With $\vec\varphi$ chosen carefully, eg. $\left.\vec{\varphi}_2\right|_{\gamma=1}=(2\pi/3,-2\pi/3)$ or $\left.\vec{\varphi}_4\right|_{\gamma=1}=\left(\cos^{-1}\left(\frac{\sqrt{10}-6}{4}\right),\cos^{-1}\left(1-\sqrt{\frac{5}{8}}\right),...\right)$, higher order constraints up to $\Phi^n_L(\vec\varphi)=f^n_L(\gamma)$ are also satisfied, thus producing rotations correct to $\mathcal{O}(\epsilon^n)$.}
\end{figure}

We now satisfy the constraints $\mathcal{B}_{n,L}^{\text{PD}}$ and $\mathcal{B}_{n,L}^{\text{AP}}$ with sequences of length exactly $L=2n$, using three different methods -- analytical, perturbative, and numerical -- and achieving the linear lower bound for $2\pi$-pulse sequences. Our solutions are non-recursive; a lower order sequence never appears as part of an order $n$ sequence. Table \ref{sequence_table} summarizes our results labeled by PD$n$ (AP$n$) for the palindromes (antipalindromes), as well as ``Tower of Power" (ToP$n$) sequences, a name inspired by their visual appearance in Fig.~\ref{APLiePlot}, which are special AP$n$ sequences essential to our perturbative proof for that length-optimal arbitrary $n$ corrections for non-trivial $\gamma$ exist.

Subsections \ref{Closed form solutions}, \ref{Perturbative solutions}, and \ref{Numerical solutions} detail respectively the analytical, perturbative, and numerical solution methods and the corresponding results.

\subsection{Closed form solutions}
\label{Closed form solutions}

We obtain closed-form solutions to $\mathcal{B}_{n,2n}^{\text{AP}}$ and $\mathcal{B}_{n,2n}^{\text{PD}}$ for $n\le 3(4)$, presented in Table \ref{table_closed_form}, by the method of Gr\"{o}bner bases \cite{Cox2007}, which we describe here. The sequences AP$2$, AP$3$, and PD$4$ are original whereas AP$1$ and PD$2$ recover SK$1$ and PB$_1$ of Brown et al. \cite{Brown2004} and Wimperis \cite{Wimperis1994} respectively.

\renewcommand{\arraystretch}{1.4}
\begin{table*}[ht]
\begin{tabular}{|c|c|l|}
\hline
Sequence & n & Phase angle solutions $\varphi_k=2 \tan^{-1}(t_k)$\\
\hline
 AP$1$ & 1 & $\begin{array} {rl}
    h(t_1)=&(2+\gamma)-(2-\gamma)t_1^2.
    \end{array}$\\
\hline
 AP$2$ & 2 & $\begin{array} {rl}
    h(t_1)=&(\gamma +2)^2 (\gamma +4)+2 \gamma  \left(\gamma ^2+4\right) t_1^2+(\gamma -4) (\gamma -2)^2 t_1^4,\\
    t_2=&t_1\frac{2-\gamma }{2+\gamma}.
    \end{array}$\\
\hline
 PD$2$ & 2 & $\begin{array} {rl}
    h(t_1)=&(4+\gamma)-(4-\gamma)t_1^2,\\
    t_2=&-t_1.
    \end{array}$\\
 \hline
    NS$2$ & 2 & $\begin{array} {rl}
    h(t_1,t_4)=&(\gamma+2)^2(\gamma+4)+\gamma(\gamma^2-4)(t_1^2+t_4^2)-16\gamma t_1 t_4+(\gamma-4)(\gamma-2)^2 t_1^2 t_4^2,\\
    t_{2(3)}=&t_{1(4)}\frac{(\gamma -4) (\gamma -2)-8 (1-\gamma)\left(1+t_{4(1)}^2\right)^{-1}-8\left(1+t_{1(4)}^2\right)^{-1}}{(4-\gamma ) (\gamma +2)-8 (\gamma
   +1)\left(1+t_{4(1)}^2\right)^{-1}-8\left(1+t_{1(4)}^2\right)^{-1}}.
    \end{array}$\\
 \hline
 AP$3$ & 3 & $\begin{array} {rl}
    h(t_1)=&a_1(\gamma) + 4 a_2(\gamma)t_1^2 + 6 a_3(\gamma)t_1^4 + 4 a_2(-\gamma)t_1^6 + a_1(-\gamma)t_1^8,\quad\text{where}\\
    a_1(\gamma)&=(\gamma -2) (\gamma +2)^3 (\gamma +4)^2 (\gamma +6)^2,\\
    a_2(\gamma)&=(\gamma +2) (\gamma +4) (\gamma +6) \left(\gamma ^5-32 \gamma ^3+96 \gamma ^2+256 \gamma +192\right),\\
    a_3(\gamma)&=\gamma ^8-60 \gamma ^6+816 \gamma ^4+9152 \gamma ^2-9216,\\
    t_2^2=&-\frac{96 \gamma +16 \left(\gamma ^2-3 \gamma +2\right) \left(t_1^2+1\right)+(\gamma -4) (\gamma -2) (\gamma +2) \left(t_1^2+1\right){}^2}{96 \gamma +16 \left(\gamma ^2-9 \gamma +2\right) \left(t_1^2+1\right)+(\gamma -6) (\gamma -4) (\gamma -2) \left(t_1^2+1\right){}^2},\\
    t_3^2=&-\frac{\gamma +2}{\gamma - 2}\frac{16 (\gamma +1)+(\gamma -4) (\gamma -2) \left(t_1^2+1\right)}{16 (\gamma -1)+(\gamma -6) (\gamma -4) \left(t_1^2+1\right)},\quad\text{where signs of }t_{2,3}\text{ chosen to satisfy}\\
    0&=\frac{t_1}{s_1}\left(\frac{2-s_1}{s_1}+2\frac{2-s_2}{s_2}+2\frac{2-s_3}{s_3}\right)+\frac{t_2}{s_2}\left(\frac{2-s_2}{s_2}+2\frac{2-s_3}{s_3}\right)+\frac{t_3}{s_3}\frac{2-s_3}{s_3},\quad s_n=1+t_n^2.

         \end{array}$\\

 \hline
 PD$4$ & 4 & $\begin{array} {rl}
    h(t_1)=&a_1(-\gamma)+3 b_1(\gamma)t_1^2-3 b_1(-\gamma)t_1^4-a_1(\gamma)t_1^6,\\
   t_n=&t_1(-1)^{1+\lceil n/2\rceil} \frac{b_n(\gamma)+a_n(\gamma)t_1^2}{a_n(-\gamma)+b_n(-\gamma)t_1^2},\quad n=2,3,4,\\
   a_n(\gamma)&=(\gamma -8) (\gamma -4) (\gamma +4)\begin{cases}
    -(\gamma -8) (\gamma -4)^2 \gamma &, n =1, \\
        -3\gamma \left(\gamma ^2-2 \gamma +4\right) & , n = 2,\\
       \gamma  (\gamma +2) \left(\gamma ^2-6 \gamma -4\right) &, n= 3,\\
        \gamma -4 &, n =4
         \end{cases}\\
   b_n(\gamma)&=(8+\gamma)\begin{cases} \gamma ^6-64 \gamma ^4+128 \gamma ^3+1024 \gamma ^2+1024 \gamma -1024 &, n =1, \\
        5 \gamma ^5-10 \gamma ^4-76 \gamma ^3-64 \gamma ^2-64 \gamma +128 & , n = 2,\\
        \left(\gamma ^2-6 \gamma -4\right) \left(\gamma ^4+2 \gamma ^3-48 \gamma ^2-32 \gamma -64\right) &, n= 3,\\
        \gamma ^3+4 \gamma ^2-64 \gamma +32 &, n =4
         \end{cases}\\
    \end{array}$\\
 \hline
\end{tabular}
\caption{\label{table_closed_form} Closed-form solutions of phase angles $\varphi_k=2 \tan^{-1}(t_k)$ for pulse sequences correcting to order $\mathcal{O}(\epsilon^n)$ represented as regular chains, computed by the method of Gr\"{o}bner bases \cite{Cox2007}. Since $h(t_1)=0$ is a univariate polynomial of degree $\le 4$ in $t_1^2$, it can be solved in closed form. The other $t_k$ are obtained directly by substitution. NS$2$ generalizes AP$2$ and PD$2$ with one free parameter $t_4$ which should be fixed before solving.}
\end{table*}

The key insight in solving the transcendental constraints $\mathcal{B}_{n,L}^{\text{AP}}$ and $\mathcal{B}_{n,L}^{\text{PD}}$ with $\gamma$ as a free parameter is that \emph{any} $\mathcal{B}_{n,L}$ is equivalent to systems of multivariate polynomial equations $\mathcal{F}$, for which powerful algorithmic methods of solution are known. This equivalence can be seen by introducing the Weierstrass substitution $\tan(\varphi_k/2)=t_k$. Any $\mathcal{F}$ with variables $t_1,...,t_n\in \mathbb{C}$ that has a finite number of zeroes is zero-dimensional and has solutions that can always be represented in the form of a regular chain, that is, a finite triangular system of polynomials $\{h_1(t_1), h_2(t_1,t_2),\cdots,h_n(t_1,...,t_n)\}$ obtained by taking appropriate linear combinations of elements of $\mathcal{F}$. Regular chains are easy to solve as the first equation $h_1$ is a univariate polynomial in $t_1$ whose zeroes can then be substituted into $h_2$, thus converting it into a univariate polynomial in $t_2$. Through recursive substitution, all $t_k$ can obtained in a straightforward manner.

Divining these appropriate linear combinations appears to be a formidable task, but surprisingly, they can be deterministically computed by applying algorithms such as Buchberger's algorithm \cite{Buchberger2006} for computing the Gr\"{o}bner basis \cite{Cox2007} $\mathcal{G}$ of $\mathcal{F}$. The basis $\mathcal{G}$ is another system of polynomial equations that shares the same zeroes as $\mathcal{F}$, in addition to certain desirable algebraic properties. For example, $\mathcal{G}$ can readily decide the existence, number of, and location of complex zeroes \cite{Buchberger1998}, and by choosing a lexicographic \emph{term order}, $\mathcal{G}$ is itself a regular chain \cite{Cox2007}. The algorithm generalizes Gaussian elimination for systems of linear equations and finding the greatest common divisor of univariate polynomial equations to systems multivariate polynomial equations: the reader is referred to excellent resources for more information \cite{Sturmfels2005,Sturmfels2002,Cox2007}. In the Appendix, we also present a brief overview of Gr\"{o}bner bases and Buchberger's algorithm for calculating them, including hand-worked examples for AP$1$ and PD$2$, the results of which are part of Table $\ref{table_closed_form}$.

The regular chains for the remaining sequences AP$2$, AP$3$, and PD$4$ solved in Table \ref{table_closed_form} can be computed by optimized variants of Buchberger's algorithm \cite{Cox2007} in Mathematica. In each case, closed-form is achieved since $h(t_1)$ is a univariate polynomial of at most quartic degree in $t_1^2$, and the remaining variables $t_{2,..,n}$ are then given as functions of only $t_1$. Only the real solutions, which exist for $|\gamma|\le2\lfloor n/2\rfloor+2$, are physically meaningful. The utility of Gr\"{o}bner bases for short sequences is clear as is it highly unlikely that these solutions could have been arrived at by hand.

As a curiosity, we also present in Table \ref{table_closed_form} a closed-form solution for $\mathcal{B}^{2\pi}_{2,4}$, where \emph{no} symmetry has been applied to the four pulse sequence, denoted NS$2$. NS$2$ has one free parameter in the phase angles, which we arbitrarily choose to be $t_4$. By fixing $t_4$ and solving for the remaining phase angles, one finds that NS$2$ continuously deforms between PD$2$ and AP$2$ and hence generalizes them.

Could one solve $\mathcal{B}^{2\pi}_{n,2n}$ for arbitrary $n$ by this method? Any arbitrary system of multivariate polynomials is guaranteed to have a Gr\"{o}bner basis that can always computed in a finite number of steps by Buchberger's algorithm \cite{Buchberger2006}. Thus, complex solutions to zero-dimensional $\mathcal{B}^{2\pi}_{n,2n}$ with the same number of equations as free parameters $\varphi_k$ can always be found by this method in principle. However, the worst-case time complexity of computing $\mathcal{G}$ for a system of $n$ variables and total degree $d$ scales as $\mathcal{O}(d^{2^{n}})$ \cite{Dube1990} and rapidly becomes infeasible. Of greater concern, there is no guarantee that such solutions are real, representing physical phases $\phi_k$.

We now prove that there exists real solutions to $\mathcal{B}^{2\pi}_{n,2n}$ over a continuous range of $\gamma$ for arbitrary $n$, and show how this leads to an efficient constructive procedure for computing arbitrary angle sequences.

\subsection{Perturbative solutions}
\label{Perturbative solutions}

We may solve $\mathcal{B}_{n,2n}^{\text{AP}}$ and $\mathcal{B}_{n,2n}^{\text{PD}}$ perturbatively. A well-known theorem of square Jacobian matrices states that any arbitrary function $\mathcal{H}(\vec{\varphi}):\mathbb{R}^n\rightarrow \mathbb{R}^n$ is locally invertible, or analytical, in the neighbourhood about some point $\vec{\varphi}_0$ if and only if the determinant $\det{(J)}$ of its Jacobian matrix $J_{jk}=\partial_{\varphi_k}\mathcal{H}_j|_{\vec{\varphi}=\vec{\varphi}_0}$ is non-zero. Thus, setting $\mathcal{H}=\mathcal{B}_{n,L}$, this theorem says that one may always construct a perturbative expansion for $\vec{\varphi}$ over a continuous range $\gamma$ about $\gamma_0$ given a valid starting point $(\vec{\varphi}_0,\gamma_0)$ satisfying $\mathcal{B}_{n,L}$ if and only if $\det{J}\neq0$. So long as the Jacobian remains non-zero, one may extend such a solution beyond its neighbourhood by analytic continuation.

However, for arbitrary $n$, what are these valid initial points $(\vec{\varphi}_0,\gamma_0)$? As we do not {\it a priori} know of solutions to $\mathcal{B}_{n,L}$ for arbitrary $\gamma$, such points must be found at some $\gamma$ where the problem simplifies. Even then, the problem is non-trivial: for example, imposing phase angle symmetries forces $\varphi_k=m\frac{\pi}{2},\,m\in\mathbb{Z}$ at $\gamma=0$, but one can readily verify that many such solutions of this form to Eq. \ref{phase_sum=ffunction} suffer from $\det{J}=0$.
Using the closed-form solutions in Table \ref{table_closed_form}, one finds at $\gamma=0$ that while the Jacobian of the PD$2,4$ sequences is zero, the AP$1,3$ sequences each have a solution with non-zero Jacobian wherein $\varphi_k=\pi/2,\text{ for}\, k\le n$. We now prove that this generalizes to arbitrary $n$, resulting in the special class of $\text{ToP}n$ antipalindrome sequences with initial values
\begin{equation}
\label{ToPn_Initial_values}
\left.\vec\varphi^{\text{ToP}}_n\right|_{\gamma=2b}=\begin{cases}\pi,&1\le k\le b\\\pi/2,&b<k\le n,\end{cases}
\end{equation}
for $b=0,1,\dots\lfloor n/2\rfloor$. Hence, non-trivial real solutions to $\mathcal{B}_{n,2n}^{2\pi}$ exist for arbitrary $n$.
\subsubsection{ToP$n$ is analytical at $\gamma=0$ $\forall n$}

We first transform the function mapping for ToP$n$: $\mathcal{H}(\vec{\varphi})_j={\big\{\begin{smallmatrix}\Rep\Phi^j_{2n}(\vec\varphi)&\text{,  }j\text{ odd}\\\Imp\Phi^j_{2n}(\vec\varphi)&\text{,  }j\text{ even },\end{smallmatrix}}\rightarrow{\big\{\begin{smallmatrix}\Phi^j_{2n}(\vec\varphi)&\text{,  }j\text{ odd}\\-i\Phi^j_{2n}(\vec\varphi)&\text{,  }j\text{ even }.\end{smallmatrix}}$ This does not affect the magnitude of its Jacobian $J^n$ as $\Rep\Phi^j_L(\vec\varphi)=\Phi^j_L(\vec\varphi)$ for odd $j$ due to antipalindromic symmetry, while for even $j$ the real part of Eq. \ref{phase_sum=ffunction} is automatically satisfied due to unitarity (see Eq. \ref{leading_error}).

The $\gamma=0$ solution to ToP$n$ has a simple form $\varphi^{\text{ToP}}_k=\pi/2,\text{ for}\, k\le n$. With this solution, a straightforward, if tedious, manipulation of the phase sums shows that elements of Jacobian matrix satisfy the recurrence $J^n_{j+1,k+1}=J^n_{j+1,k}+J^n_{j,k+1}+J^n_{jk}$. The solution to this recurrence is best seen from a combinatorial standpoint. Consider the related puzzle --- you begin at the top left corner (1,1) of an $s\times k$ checkerboard and would like to reach the position $(s,k)$, the lower right corner. You may move only south, southeast, or east at any given time, enforcing the recursion. If, additionally, your first move cannot be south, how many paths exist that achieve your goal? The solution is
\begin{equation}
W_{sk}=\sum_{r=0}^{s-1}\binom{k-1}{s-r-1}\binom{k+r-2}{r},
\end{equation}
since you may take any number of southerly steps $r$. If your first move is not restricted, the number of paths is $D_{sk}=\sum_{p=1}^sW_{pk}=\sum_{r=0}^{s-1}\binom{k-1}{s-r-1}\binom{k+r-1}{r}$. For later use, define an $n\times n$ matrix $D^n$ with the elements $D^n_{jk}=D_{jk}$.

We will now express $J^n_{jk}$ in terms of the leftmost column $J^n_{j1}$. This is an extension of the path counting problem, in which we may begin our walk to $(j,k)$ from any leftmost location. Therefore,
\begin{align}
J^n_{jk}&=\sum_{s=1}^{j}T^n_{j-s+1,1}W_{sk}
\end{align}
Now notice that the determinant of $J^n$ does not depend upon the leftmost column. Since $J^n_{jk}=J_{j1}W_{0k}+\dots+J_{11}W_{jk}$ and $J_{11}=-2\neq0$ we can always subtract multiples of rows of $J^n$ to obtain $-2D^n$. Thus, $\det(J^n)=(-2)^n\det(D^n)$.

We have reduced the problem to finding the determinant of $D^n$. We claim that $D^n$ has LU-decomposition
\begin{equation}\label{LU-decomp}
D^n_{jk}=\sum_{h=1}^n\binom{j-1}{h-1}U^n_{hk},\quad U^n_{hk}\equiv2^{h-1}\binom{k-1}{h-1}.
\end{equation}
This means that $D^n_{jk}$ is the binomial transform of the Chebyshev triangle $U^n_{hk}$. This is proved by looking at the generating functions of $D^n$ and $U^n$, namely
\begin{align}\label{gen_D}
\mathcal{D}(y,z)&=\sum_{j,k=1}^\infty D^n_{jk}y^{j-1}z^{k-1}=\frac{1}{1-(y+yz+z)},\\\label{gen_U}
\mathcal{U}(y,z)&=\sum_{j,k=1}^\infty U^n_{jk}y^{j-1}z^{k-1}=\frac{1}{1-(1+2y)z}.
\end{align}
These are related by $\mathcal{D}(y,z)=\frac{1}{1-y}\mathcal{U}\left(\frac{y}{1-y},z\right)$, which implies the binomial transform in Eq.~\eqref{LU-decomp}. With the LU-decomposition, one can immediately see that $\det(D^n)=\det(U^n)=2^{n(n-1)/2}$, we have $\det(J^n)=(-1)^n 2^{n(n+1)/2}\neq0$. This concludes the proof that ToP$n$ sequences exist for a continuous range of small target angles $\gamma$ within the neighbourhood of $\gamma_0$ all $n$.

\begin{figure}
\includegraphics[width=\columnwidth]{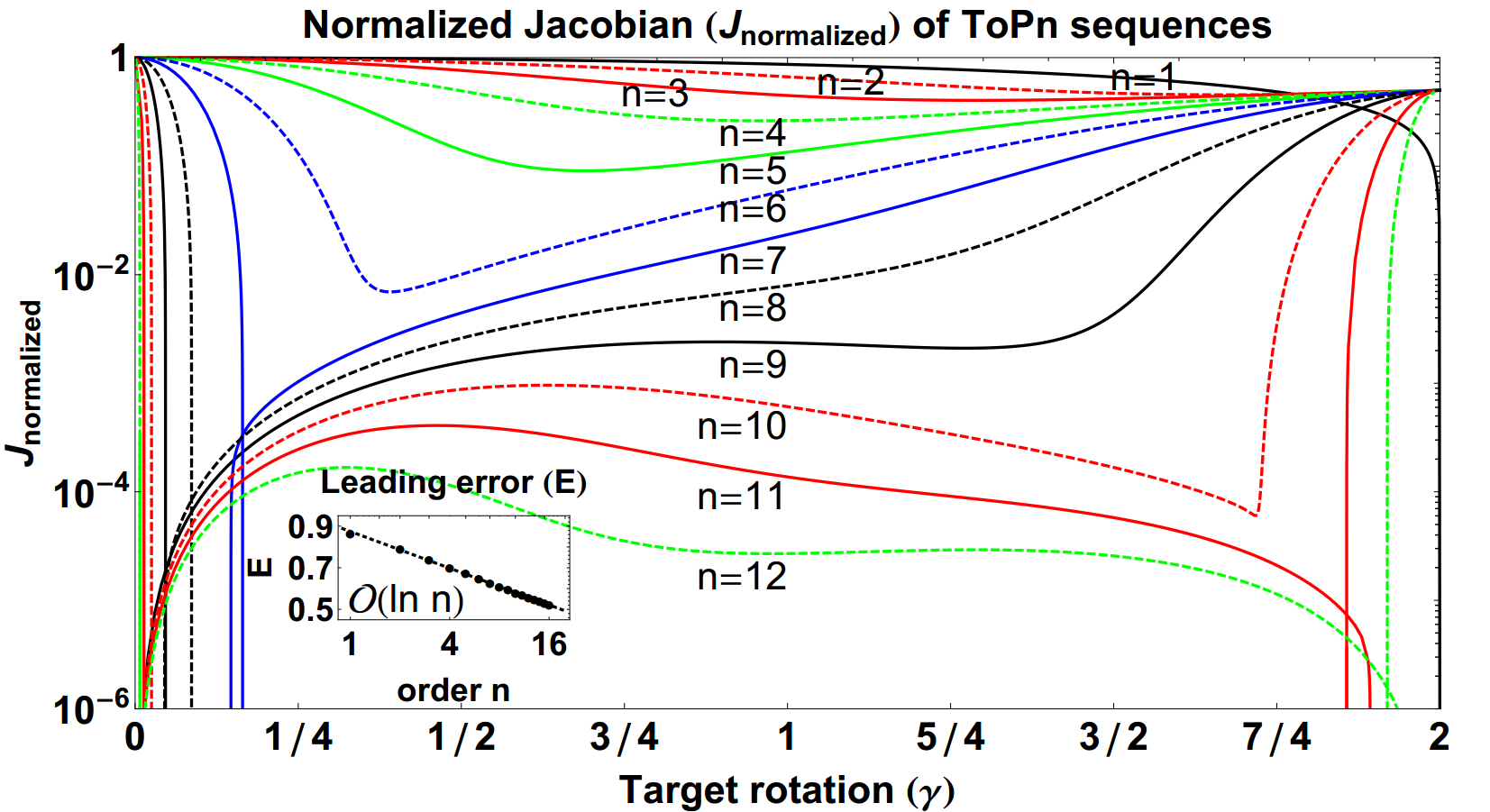}
\caption{\label{detplot} ToP$n$ Jacobian $J_{normalized}$ normalized to $1$ at $\gamma=0$ obtained by analytic continuation from zero-order solutions at $\gamma=0,1,2$ for $n\le12$. Analytic continuation to arbitrary angles from $\gamma=0$ is possible for $n<7$. From $7 \le n<11$, covering all $\gamma$ requires continuation from $\gamma=2$ as well. For $11\le n$, continuation from $\gamma=1$ is also necessary. The inset plots the leading error $|f_{2n}^{n+1}-\Phi_{2n}^{n+1}|$ of ToP$n$ at $\gamma=1$ up to $n=16$ together with the best-fit (dotted) $E=-\mathcal{O}(ln(n))$.}
\end{figure}

\subsection{Numerical solutions}
\label{Numerical solutions}

Our demonstrations of real, arbitrary angle ($\gamma$) solutions for small order ($n$) and real, arbitrary order solutions for a continuous range of small angles inspires confidence that real solutions for larger $\gamma$ at arbitrary $n$ can always be found. Although proving this notion is difficult, the zeroth-order analytically continuable solutions provide, in principle, a means of obtaining arbitrary $\gamma$, arbitrary $n$ sequences that are exponentially more efficient than a brute force search for solutions to Eq.~\eqref{phase_sum=ffunction}. As long as a sequence for some $\gamma$ has a non-zero Jacobian, its phase angles may be continuously deformed into another solution to Eq.~\eqref{phase_sum=ffunction} in the neighbourhood of $\gamma$.

We present the results of this procedure for ToP$n$ and PD$n$ to obtain real optimal length solutions over $\gamma\in[0,2]$, and provide for the convenience of the reader some solutions derived in this manner at common values of $\gamma=\{1,\frac{1}{2},\frac{1}{4}\}$ up to $n=12$ in Table.~\ref{numerical_varphi_table}.

We provide details of the continuations for ToP$n$ and PD$n$ in subsections \ref{Analytic continuation of ToP$n$} and \ref{Analytic continuation of PD$n$}. We know more exotic solutions exist, too, and in section \ref{Brute-force numerical search} we provide the results of brute force numerical solutions to Eq.~\eqref{phase_sum=ffunction}.

\renewcommand{\arraystretch}{0.85}
\begin{table*}[t,p]
\tiny
\begin{center}
\begin{tabular}{|c|| c|c|c   ||c|cc|c ||c|ccc|c|}\hline
 $\gamma$
 &$n=1$ & $\varphi _1$ & $\kappa$
 &$n=2$ & $\varphi _1$ & $\varphi _2$ & $\kappa$
 &$n=3$ & $\varphi _1$ & $\varphi _2$ & $\varphi _3$ & $\kappa$ \\
\hline
\multirow{2}{*}{1}
&\multirow{2}{*}{\bf AP1$_1$} & \multirow{2}{*}{2.09440} & \multirow{2}{*}{1.31607}
&PD2$_1$ & 1.82348 & -1.82348 & 1.16499
&AP3$_1$ & 0.74570 & -2.11099 & 2.37504 & 1.10279\\
&
&&
&\bf AP2$_1$ & 2.35949 & 1.35980 & 1.16499
&\bf AP3$_2$ & 2.51806 & 1.15532 & 1.66273 & 1.10279\\
\hline\hline
\multirow{2}{*}{$\frac{1}{2}$}
&\multirow{2}{*}{\bf AP1$_1$} & \multirow{2}{*}{1.82348} & \multirow{2}{*}{0.98400}
&PD2$_1$ & 1.69612 & -1.69612 & 0.88856
&AP3$_1$ & 0.87848 & -1.93555 & 2.13129 & 0.93360 \\
&
&&
&\bf AP2$_1$ & 1.95071 & 1.44966 & 1.05957
&\bf AP3$_2$ & 2.03611 & 1.29441 & 1.64504 & 1.12057 \\
\hline\hline
\multirow{2}{*}{$\frac{1}{4}$}
&\multirow{2}{*}{\bf AP1$_1$} & \multirow{2}{*}{1.69612} & \multirow{2}{*}{0.70433}
&PD2$_1$ & 1.63334 & -1.63334 & 0.69647
&AP3$_1$ & 0.98173 & -1.85668 & 1.98076 & 0.77869 \\
&
&&
&\bf AP2$_1$ & 1.75891 & 1.50875 & 0.86557
&\bf AP3$_2$ & 1.80090 & 1.42667 & 1.61136 & 0.98560\\ \hline
\end{tabular}
\begin{tabular}{|c || c|cccc|c || c|ccccc| c|}\hline
$\gamma$
&$n=4$ & $\varphi _1$ & $\varphi _2$ & $\varphi _3$ & $\varphi _4$ & $\kappa$
&$n=5$ & $\varphi _1$ & $\varphi _2$ & $\varphi _3$ & $\varphi _4$ & $\varphi _5$ & $\kappa$ \\
\hline
\multirow{4}{*}{$1$}
&PD4$_1$ & 2.26950 & -1.76948 & -0.80579 & 1.93044 & 1.07009
&AP5$_1$ & 2.30757 & -2.57163 & 1.03434 & -0.26267 & 2.05214 & 1.05043 \\

&PD4$_2$ & 1.38777 & -0.59527 & 2.74476 & -2.19903 & 1.07009
&AP5$_2$ & 0.44569 & -1.40804 & 1.55593 & -2.45457 & 2.50831 & 1.05043 \\

&AP4$_1$ & 1.64767 & -1.97451 & 2.92461 & 0.32956 & 1.07009
&AP5$_3$ & 2.19409 & 0.13026 & -2.09113 & 1.82984 & 1.72591 & 1.05043 \\

&\bf AP4$_2$ & 2.62323 & 0.99049 & 1.79394 & 1.52913 & 1.07009
&\bf AP5$_4$ & 2.69800 & 0.86163 & 1.92713 & 1.45034 & 1.59011 & 1.05043\\
\hline\hline
\multirow{4}{*}{$\frac{1}{2}$}
&PD4$_1$ & 2.30661 & -1.30540 & -0.47998 & 2.38635 & 0.88941
&AP5$_1$ & 1.86484 & -2.24227 & 1.42219 & -0.43481 & 1.97474 & 0.91458 \\

&PD4$_2$ & 1.47070 & 0.11256 & -2.96678 & -1.93782 & 0.89673
&AP5$_2$ & 0.60281 & -1.44347 & 1.45031 & -2.28880 & 2.29567 & 0.93241 \\

&AP4$_1$ & 1.05532 & -2.36238 & 3.06746 & 0.26240 & 0.90343
&AP5$_3$ & 1.78432 & 0.205507 & -2.21897 & 1.86132 & 1.69801 & 1.05179 \\

&\bf AP4$_2$ & 2.10426 & 1.11746 & 1.80109 & 1.52196 & 1.15341
&\bf AP5$_4$ & 2.17223 & 0.89078 & 2.05179 & 1.39042 & 1.60052 & 1.13467\\
\hline\hline
\multirow{4}{*}{$\frac{1}{4}$}
&PD4$_1$ & 2.45079 & -0.96051 & -0.28079 & 2.66423 & 0.76705
&AP5$_1$ & 1.56763 & -2.19365 & 1.57024 & -0.61251 & 1.94290 & 0.80141 \\

&PD4$_2$ & 1.51926 & 0.73603 & -2.38182 & -1.76467 & 0.77110
&AP5$_2$ & 0.73268 & -1.46554 & 1.41033 & -2.18996 & 2.15661 & 0.82226 \\

&AP4$_1$ & 0.68460 & -2.64974 & 3.11442 & 0.19308 & 0.77643
&AP5$_3$ & 1.62724 & 0.36022 & -2.23779 & 1.88279 & 1.64971 & 0.95228 \\

&\bf AP4$_2$ & 1.83302 & 1.33340 & 1.70166 & 1.54125 & 1.07442
&\bf AP5$_4$ & 1.86049 & 1.22698 & 1.85066 & 1.44825 & 1.59325 & 1.13458\\
\hline
\end{tabular}
\begin{tabular}{|c || c|cccccc|c || c|ccccccc| c|}\hline
$\gamma$
&$n=6$ & $\varphi _1$ & $\varphi _2$ & $\varphi _3$ & $\varphi _4$ & $\varphi _5$ & $\varphi _6$ & $\kappa$
&$n=7$ & $\varphi _1$ & $\varphi _2$ & $\varphi _3$ & $\varphi _4$ & $\varphi _5$ & $\varphi _6$ & $\varphi _7$ & $\kappa$ \\
\hline
\multirow{8}{*}{$1$}
&PD6$_1$ & 2.48390 & -1.63561 & -0.23686 & 2.03217 & 2.74686 & -0.71914 & 1.03757
&AP7$_1$ & 2.13314 & 1.81689 & -1.08618 & -1.35267 & 2.63930 & 0.24057 & 2.09463 & 1.02864 \\

&PD6$_2$ & 2.24222 & -2.44350 & 1.64224 & -1.08266 & -1.76926 & 0.81242 & 1.03757
&AP7$_2$ & 0.31786 & -0.94135 & 1.49035 & -1.44478 & 2.14820 & -2.58143 & 2.58988 & 1.02864 \\

&PD6$_3$ & 1.17370 & -0.19700 & 2.33177 & -0.99925 & 3.05926 & -2.37762 & 1.03757
&AP7$_3$ & 1.53566 & -1.94880 & -3.10873 & 1.06287 & -1.02408 & 0.63518 & -2.93056 & 1.02864 \\

&PD6$_4$ & 0.38266 & -3.00356 & -2.24117 & 2.23067 & -1.43621 & 0.84607 & 1.03757
&\bf AP7$_4$ & 2.79719 & 0.67921 & 2.15156 & 1.25324 & 1.70211 & 1.53591 & 1.57511 & 1.02864 \\

&AP6$_1$ & 2.54809 & 2.02296 & -0.58625 & 0.73627 & 2.99308 & 1.38890 & 1.03757
&AP7$_5$ & 2.67233 & 1.22852 & -0.41410 & 2.09290 & 3.01643 & 0.79095 & 1.64235 & 1.02864 \\

&AP6$_2$ & 1.92279 & -3.02711 & -0.18830 & -1.61442 & 2.05848 & 1.19384 & 1.03757
&AP7$_6$ & 2.38920 & 2.56740 & -1.87487 & 1.51487 & -0.58086 & 0.89490 & 1.71973 & 1.02864 \\

&AP6$_3$ & 0.74467 & -2.15643 & 2.73082 & 2.72326 & -0.85131 & 1.05986 & 1.03757
&AP7$_7$ & 1.90502 & -0.25506 & -0.74525 & 2.18555 & -2.05722 & 2.23559 & 1.78586 & 1.02864 \\

&\bf AP6$_4$ & 2.75387 & 0.76025 & 2.04746 & 1.35335 & 1.63574 & 1.56171 & 1.03757
&AP7$_8$ & 2.56602 & 0.68912 & 0.58385 & -2.14580 & 2.36676 & 1.54736 & 1.60313 & 1.02864\\
\hline\hline
\multirow{8}{*}{$\frac{1}{2}$}
&PD6$_1$ & 2.71338 & -0.89153 & 0.34947 & 2.88508 & -2.77240 & 0.12790 & 0.89347
&AP7$_1$ & 1.71155 & 2.56416 & -0.40810 & -1.38344 & 2.74047 & 0.11831 & 2.03574 & 0.92275 \\

&PD6$_2$ & 2.40846 & -1.52806 & 3.02225 & 0.14373 & -1.19151 & 1.48740 & 0.89824
&AP7$_2$ & 0.46860 & -1.07834 & 1.37212 & -1.48204 & 1.99013 & -2.42873 & 2.39994 & 0.93660 \\

&PD6$_3$ & 1.35661 & 0.50760 & 2.84949 & -0.34944 & -2.58938 & -2.05672 & 0.90808
&AP7$_3$ & 0.62501 & -2.74449 & -2.77743 & 1.15902 & -0.93150 & 0.79294 & -2.69744 & 0.94367 \\

&PD6$_4$ & 0.34769 & -2.51801 & 2.11029 & -1.90548 & 1.90034 & -0.66420 & 0.91063
&\bf AP7$_4$ & 2.33062 & 0.29389 & 2.38838 & 1.28505 & 1.66008 & 1.54990 & 1.57324 & 0.94652 \\

&AP6$_1$ & 2.11291 & 2.26524 & -0.55309 & 0.48262 & 2.87662 & 1.43607 & 0.95146
&AP7$_5$ & 2.20315 & 1.13325 & -0.64167 & 2.00550 & 2.99463 & 0.94849 & 1.62714 & 1.01054 \\

&AP6$_2$ & 1.56304 & 2.92131 & -0.50059 & -1.71787 & 1.85652 & 1.29826 & 0.97397
&AP7$_6$ & 1.92827 & 2.68589 & -1.88707 & 1.50932 & -0.51984 & 1.04819 & 1.69068 & 1.01961 \\

&AP6$_3$ & 0.97792 & -1.74876 & 2.49396 & 2.83905 & -0.72251 & 1.19155 & 0.98162
&AP7$_7$ & 1.58492 & -0.10094 & -0.98974 & 1.95733 & -2.24104 & 2.21759 & 1.75068 & 1.02465 \\

&\bf AP6$_4$ & 2.26941 & 0.51330 & 2.35282 & 1.27588 & 1.64389 & 1.56168 & 0.99856
&AP7$_8$ & 2.05179 & 0.74564 & 0.44549 & -2.29705 & 2.45454 & 1.51269 & 1.61609 & 1.09307\\
\hline\hline
\multirow{8}{*}{$\frac{1}{4}$}
&PD6$_1$ & 3.12733 & -0.27154 & 0.71713 & -2.88348 & -2.34449 & 0.60325 & 0.80563
&AP7$_1$ & 1.44528 & 3.12551 & 0.18508 & -1.47200 & 2.68933 & -0.05344 & 2.01638 & 0.83378 \\

&PD6$_2$ & 2.66911 & -0.96400 & -2.59675 & 0.76929 & -0.86098 & 1.89620 & 0.80770
&AP7$_2$ & 0.59981 & -1.15892 & 1.31704 & -1.52059 & 1.89255 & -2.33236 & 2.27275 & 0.85026 \\

&PD6$_3$ & 1.46125 & 1.00480 & 2.91140 & -0.24281 & -2.10817 & -1.82942 & 0.81322
&AP7$_3$ & 0.10274 & 3.04219 & -2.58618 & 1.23233 & -0.92136 & 0.93214 & -2.51134 & 0.85965 \\

&PD6$_4$ & 0.61658 & -2.23483 & 2.05399 & -1.70970 & 2.20472 & -0.51392 & 0.81490
&\bf AP7$_4$ & 1.97582 & 0.44006 & 2.74911 & 1.20237 & 1.65775 & 1.55408 & 1.57254 & 0.88415 \\

&AP6$_1$ & 1.82740 & 2.77886 & -0.11820 & 0.33508 & 2.95974 & 1.46074 & 0.84058
&AP7$_5$ & 1.90819 & 1.35140 & -0.95152 & 1.65205 & -3.07859 & 1.01561 & 1.61079 & 0.89826 \\

&AP6$_2$ & 1.43968 & 2.68323 & -0.67496 & -1.73584 & 1.74153 & 1.37503 & 0.87521
&AP7$_6$ & 1.70670 & 2.60476 & -1.80222 & 1.60237 & -0.56759 & 1.19451 & 1.65081 & 0.93279 \\

&AP6$_3$ & 1.09750 & -1.53786 & 2.38588 & 2.88953 & -0.61467 & 1.30211 & 0.88747
&AP7$_7$ & 1.48964 & 0.08106 & -1.25334 & 1.70494 & -2.32236 & 2.18736 & 1.69126 & 0.94539 \\

&\bf AP6$_4$ & 1.88994 & 1.07769 & 2.11382 & 1.23820 & 1.68167 & 1.55453 & 1.14696
&AP7$_8$ & 1.78462 & 1.06489 & 0.28111 & -2.37747 & 2.24540 & 1.54878 & 1.60439 & 1.08061\\
\hline
\end{tabular}
\begin{tabular}{|c || c|cccccccc|c |}\hline
$\gamma$
&$n=8$ & $\varphi _1$ & $\varphi _2$ & $\varphi _3$ & $\varphi _4$ & $\varphi _5$ & $\varphi _6$ & $\varphi _7$ & $\varphi _8$ &$\kappa$ \\
\hline
\multirow{16}{*}{$1$}
&PD8$_1$ & 2.61153 & -1.49208 & 0.15281 & 2.20741 & -2.95859 & -0.58046 & -0.00546 & 2.86733 & 1.02216 \\
&PD8$_2$ & 2.50835 & -1.80323 & -1.12170 & -3.08779 & 0.28846 & 1.09114 & 1.27599 & -2.22667 & 1.02216 \\
&PD8$_3$ & 2.18294 & -2.97594 & 1.35331 & -2.84491 & 0.16565 & -1.09595 & -2.25027 & 0.16552 & 1.02216 \\
&PD8$_4$ & 1.82404 & -0.10574 & -2.98208 & -2.13737 & -1.50007 & 3.09155 & 0.11098 & 1.35471 & 1.02216 \\
&PD8$_5$ & 1.75477 & -2.32150 & 0.12777 & -1.18259 & 2.40815 & -1.93275 & 0.30847 & 2.60823 & 1.02216 \\
&PD8$_6$ & 1.03767 & -0.07640 & 1.75380 & -1.00830 & 2.33601 & -1.44271 & -3.09970 & -2.48577 & 1.02216 \\
&PD8$_7$ & 1.01659 & -2.52557 & 2.89648 & 0.33374 & -1.50338 & 2.19623 & -2.29474 & 0.09358 & 1.02216 \\
&PD8$_8$ & 1.74334 & -1.79167 & -2.39342 & -0.72709 & 2.05671 & 0.82258 & 2.48855 & -1.09561 & 1.02216 \\
&AP8$_1$ & 0.83305 & -1.73272 & 1.95053 & -2.32641 & 2.89702 & -2.63823 & 0.38560 & 0.26672 & 1.02216 \\
&AP8$_2$ & 1.61633 & -2.51472 & -1.51594 & 1.42758 & 0.18349 & -2.50110 & 2.76220 & 0.43869 & 1.02216 \\
&AP8$_3$ & 2.27849 & -2.00209 & 1.75303 & -0.08067 & 0.27972 & 2.94649 & -2.81981 & 0.76531 & 1.02216 \\
&\bf AP8$_4$ & 2.83177 & 0.61330 & 2.24038 & 1.15822 & 1.77967 & 1.49245 & 1.58945 & 1.56873 & 1.02216 \\
&AP8$_5$ & 2.35733 & 1.05688 & -2.82410 & -0.44884 & -1.30394 & 2.16134 & 1.60198 & 1.48100 & 1.02216 \\
&AP8$_6$ & 2.13392 & -0.24881 & -2.41837 & 1.70116 & 2.70715 & -1.29687 & 1.07784 & 1.46294 & 1.02216 \\
&AP8$_7$ & 2.59142 & 1.60949 & 0.74701 & -2.11111 & -0.81043 & 1.71959 & 2.01457 & 1.51056 & 1.02216 \\
&AP8$_8$ & 2.74925 & 0.90733 & 0.84501 & -2.90598 & 2.13587 & 0.18584 & 1.93876 & 1.54157 & 1.02216 \\
\hline\hline
\multirow{16}{*}{$\frac{1}{2}$}
&PD8$_1$ & 3.09896 & -0.40393 & 1.07238 & -2.81247 & -1.81742 & 0.74796 & 1.25751 & -2.09357 & 0.89860 \\
&PD8$_2$ & 2.89767 & -0.75419 & 0.02350 & -0.44104 & 2.77895 & 2.40824 & 2.36633 & -1.07052 & 0.90078 \\
&PD8$_3$ & 2.44672 & -1.76010 & 2.76238 & -2.67594 & 0.42718 & 0.16684 & -1.48585 & 0.99119 & 0.90698 \\
&PD8$_4$ & 2.15221 & 1.77359 & -1.28384 & -1.48037 & -0.86524 & -1.76586 & 1.44250 & 2.04352 & 0.90990 \\
&PD8$_5$ & 1.79499 & -2.20241 & 0.33989 & -0.13113 & -3.09034 & -1.74831 & 0.64086 & 3.07444 & 0.91557 \\
&PD8$_6$ & 1.28080 & 0.59213 & 2.03022 & -0.76950 & 2.47232 & -1.08139 & -2.48749 & -2.13266 & 0.91767 \\
&PD8$_7$ & 0.21511 & 3.01940 & 3.12108 & 0.36417 & -1.25531 & 2.24040 & -2.60242 & -0.03202 & 0.92079 \\
&PD8$_8$ & 0.97746 & -2.40477 & -2.19087 & -1.03016 & 1.93847 & 0.88825 & 2.35917 & -1.11989 & 0.92143 \\
&AP8$_1$ & 0.61590 & -1.54138 & 1.79633 & -2.64106 & 3.03586 & -2.81243 & 0.25918 & 0.20325 & 0.92361 \\
&AP8$_2$ & 1.07682 & -2.72870 & -2.26979 & 0.70542 & 0.23122 & -2.61839 & 2.88367 & 0.36618 & 0.92461 \\
&AP8$_3$ & 1.69369 & -1.97187 & 2.20200 & -0.14932 & 0.17567 & 2.86682 & -2.78478 & 0.67753 & 0.92852 \\
&\bf AP8$_4$ & 2.36536 & 0.18413 & 2.37615 & 1.28246 & 1.68628 & 1.52971 & 1.58055 & 1.56971 & 0.95723 \\
&AP8$_5$ & 1.88735 & 1.26877 & -2.67285 & -0.31979 & -1.39755 & 2.12492 & 1.60375 & 1.47723 & 1.05962 \\
&AP8$_6$ & 1.73150 & -0.17922 & -2.55773 & 1.79669 & 2.76172 & -1.23074 & 1.08662 & 1.45651 & 1.05991 \\
&AP8$_7$ & 2.09427 & 1.83387 & 0.60461 & -2.17505 & -0.63648 & 1.77937 & 1.98533 & 1.50975 & 1.06029 \\
&AP8$_8$ & 2.24717 & 0.77780 & 1.17411 & -2.62726 & 2.07274 & 0.36619 & 1.91234 & 1.54065 & 1.07079 \\
\hline\hline
\multirow{16}{*}{$\frac{1}{4}$}
&PD8$_1$ & 2.36533 & -0.59348 & -1.59387 & 2.11324 & 1.26477 & -1.39671 & -1.84750 & 1.47308 & 0.83193 \\
&PD8$_2$ & 2.77667 & -0.07740 & -0.51182 & -0.44784 & 2.63816 & -3.06582 & -2.91175 & 0.48286 & 0.83255 \\
&PD8$_3$ & 2.78074 & -1.06236 & -2.83926 & -2.58960 & 0.52429 & 0.80444 & -1.06493 & 1.48454 & 0.83488 \\
&PD8$_4$ & 2.42390 & 2.91231 & -0.17275 & -0.98092 & -0.48142 & -0.88408 & 2.29932 & 2.48783 & 0.83624 \\
&PD8$_5$ & 1.90609 & -1.94011 & 0.59178 & 0.66409 & -2.37678 & -1.59005 & 0.85358 & -2.89166 & 0.83956 \\
&PD8$_6$ & 1.42220 & 1.05360 & 1.96382 & -0.77487 & 2.39390 & -1.16804 & -2.05033 & -1.87167 & 0.84089 \\
&PD8$_7$ & 0.15447 & -2.65067 & 3.11263 & -0.30804 & 1.12308 & -2.27416 & 2.84794 & 0.16435 & 0.84309 \\
&PD8$_8$ & 0.52355 & -2.82738 & -2.13271 & -1.37376 & 1.67229 & 0.88411 & 2.21759 & -1.19502 & 0.84360 \\
&AP8$_1$ & 0.43759 & -1.25361 & 1.97034 & -2.83601 & 3.08638 & -2.92195 & 0.17407 & 0.14459 & 0.84606 \\
&AP8$_2$ & 0.74902 & -2.91162 & -2.58867 & 0.39895 & 0.21710 & -2.70507 & 2.94458 & 0.29709 & 0.84732 \\
&AP8$_3$ & 1.25748 & -2.21044 & 2.40264 & -0.18148 & 0.13282 & 2.86776 & -2.79560 & 0.58885 & 0.85238 \\
&\bf AP8$_4$ & 1.99996 & 0.25518 & 2.71191 & 1.27163 & 1.65234 & 1.54671 & 1.57614 & 1.57022 & 0.89121 \\
&AP8$_5$ & 1.68405 & 1.39892 & -2.66779 & -0.14679 & -1.40829 & 2.13276 & 1.54453 & 1.50840 & 1.00109 \\
&AP8$_6$ & 1.58930 & 0.07907 & -2.47008 & 1.94431 & 2.74375 & -1.08079 & 1.13869 & 1.49385 & 1.00433 \\
&AP8$_7$ & 1.82114 & 1.97469 & 0.72846 & -2.07905 & -0.55151 & 1.93583 & 1.85296 & 1.53092 & 0.99405 \\
&AP8$_8$ & 1.93220 & 0.95664 & 1.13275 & -2.40924 & 2.34418 & 0.46327 & 1.81690 & 1.55146 & 0.98427 \\
\hline
\end{tabular}
\begin{tabular}{|c || c|*{12}{c}|c |}\hline
$\gamma$ & $\mathcal{S}$ & $\varphi _1$ & $\varphi _2$ & $\varphi _3$ & $\varphi _4$ & $\varphi _5$ & $\varphi _6$ & $\varphi _7$ & $\varphi _8$ & $\varphi _9$ & $\varphi _{10}$ & $\varphi _{11}$ & $\varphi _{12}$ & $\kappa$ \\
\hline
\multirow{5}{*}{$1$}
& \bf AP9 & 2.86001 & 0.558795 & 2.31608 & 1.07164 & 1.86035 & 1.43542 & 1.61698 & 1.56086 & 1.57179 & - &- & - & 1.01731\\
& PD10$_1$ & 2.6967 & -1.35876 & 0.446723 & 2.41011 & -2.50642 & -0.38506 & 0.558564 & 3.03426 & -2.76626 & 0.146207 & - & - & 1.01358 \\
& \bf AP10 & 2.88351 & 0.513054 & 2.38089 & 0.994249 & 1.93898 & 1.37051 & 1.65721 & 1.54389 & 1.57606 & 1.57032 &- & - & 1.01358 \\
& \bf AP11 & 2.90338 & 0.474161 & 2.43675 & 0.925592 & 2.01287 & 1.30273 & 1.70718 & 1.51647 & 1.58631 & 1.56802 & 1.57103 & - & 1.01065\\
& PD12$_1$ & 2.7578 & -1.2404 & 0.681115 & 2.6173 & -2.13289 & -0.159668 & 1.0323 & -3.03873 & -2.21969 & 0.335138 & 0.751611 & -2.58956 & 1.0083\\
& \bf AP12 & 2.92039 & 0.440707 & 2.48526 & 0.86478 & 2.08092 & 1.2356 & 1.76313 & 1.47942 & 1.60447 & 1.56194 & 1.57226 & 1.57068 & 1.0083\\
\hline\hline
\multirow{5}{*}{$\frac{1}{2}$}
& \bf AP9 & 2.39185 & 0.108021 & 2.35997 & 1.26948 & 1.7233 & 1.4969 & 1.59735 & 1.56486 & 1.57141 & - &- & - & 0.980115\\
& PD10$_1$ & 2.63663 & -0.293436 & -1.78826 & 2.13382 & 0.911607 & -1.45848 & -2.30168 & 1.36696 & 0.935993 & -2.02087 & - & - & 0.905375\\
& \bf AP10 & 2.88351 & 0.513054 & 2.38089 & 0.994249 & 1.93898 & 1.37051 & 1.65721 & 1.54389 & 1.57606 & 1.57032 &- & - & 1.00431\\
& \bf AP11 & 2.90338 & 0.474161 & 2.43675 & 0.925592 & 2.01287 & 1.30273 & 1.70718 & 1.51647 & 1.58631 & 1.56802 & 1.57103 & - & 1.0257\\
& PD12$_1$ & 2.7578 & -1.2404 & 0.681115 & 2.6173 & -2.13289 & -0.159668 & 1.0323 & -3.03873 & -2.21969 & 0.335138 & 0.751611 & -2.58956 & 0.913103\\
& \bf AP12 & 2.92039 & 0.440707 & 2.48526 & 0.86478 & 2.08092 & 1.2356 & 1.76313 & 1.47942 & 1.60447 & 1.56194 & 1.57226 & 1.57068 & 1.03977\\
\hline\hline
\multirow{5}{*}{$\frac{1}{4}$}
& \bf AP9 & 2.01657 & 0.132932 & 2.66164 & 1.30204 & 1.66304 & 1.53099 & 1.58483 & 1.56766 & 1.57112 & - &- & - & 0.92234\\
& PD10$_1$ & 1.17079 & -1.85566 & -2.41411 & 1.2885 & 0.292084 & -2.20261 & -2.94761 & 0.654295 & 0.26564 & -2.7124 & - & - & 0.852017\\
& \bf AP10 & 2.03052 & 0.0347255 & 2.61071 & 1.31671 & 1.68306 & 1.5056 & 1.60228 & 1.56014 & 1.573 & 1.57059 &- & - & 0.955416\\
& \bf AP11 & 2.04286 & -0.0477867 & 2.56212 & 1.32168 & 1.71095 & 1.4685 & 1.63306 & 1.54272 & 1.5795 & 1.56915 & 1.57094 & - & 0.987006\\
& PD12$_1$ & 0.379182 & -2.8705 & -3.06176 & -0.386468 & 0.589951 & 3.00952 & -2.34477 & 0.147234 & 0.8115 & -2.76097 & -2.4185 & 0.573532 & 0.867854\\
& \bf AP12 & 2.05415 & -0.119278 & 2.51646 & 1.31975 & 1.74694 & 1.41647 & 1.68347 & 1.50776 & 1.59664 & 1.5635 & 1.57206 & 1.57069 & 1.01568\\
\hline
\end{tabular}
\end{center}
\caption{Tables of phase angles for AP$n$ and PD$n$ sequences at common values of $\gamma\in\{1,\frac{1}{2},\frac{1}{4}\}$ up to $n=12$. Sequences with the same subscript are related by analytic continuation, and are sorted by their leading error $\kappa$ at $\gamma=\frac{1}{2}$. The bolded AP$n$ are ToP$n$ sequences, and all PD$n_1$ sequences are obtained by analytic continuation from $\vec{\varphi}^{\text{PD}}_{n,n/2}|_{\gamma=2}$.  Note that $\varphi^{\text{AP}}_k=-\varphi^{\text{AP}}_{L-k+1}$ and $\varphi^{\text{PD}}_k=\varphi^{\text{PD}}_{L-k+1}$. Note that for each sequence $\vec\varphi$ listed here, there is a sequence $-\vec\varphi$ with the same leading error.}
\label{numerical_varphi_table}
\end{table*}

\subsubsection{Analytic continuation of ToP$n$}
\label{Analytic continuation of ToP$n$}

We plot the Jacobian of ToP$n$ solutions obtained by analytic continuation as a function of target angle $\gamma$ in Fig.~\ref{detplot}. The zeroth-order ToP$n$ solutions in Eq. \ref{ToPn_Initial_values} can be continued from $\gamma=0$ to arbitrary $\gamma$, up to $n=7$ as $J$ is non-zero over the range $\gamma\in[0,2]$. For $8 \le n \le 10$ analytic continuation from $\gamma=2$ is required as well to cover the full range of $\gamma$ as $J=0$ at small $\gamma$ as seen in Fig.~\ref{detplot} (inset). For $11\le n$, we appear to encounter some difficulty as $J=0$ near $\gamma=0,2$. However, inspecting Fig. \ref{APLiePlot} suggests that an order $n+1$ sequence at $\gamma=1$ is in some sense approximated by appending to order $n$ sequence the phase angle $\varphi_{n+1}=\pi/2$, or $\left.\vec{\varphi}^{\text{ToP}}_{n+1}\right|_{\gamma=1}\approx\left.\vec{\varphi}^{\text{ToP}}_{n}\right|_{\gamma=1}\circ(\pi/2)$. This approximation is qualified by observing the monotonic decrease of the leading error $|f_{2n}^{n+1}-\Phi_{2n}^{n+1}|$ from Eq. \ref{leading_error} for ToP$n$ sequences at $\gamma=1$. Thus, in the limit where $n\rightarrow\infty$, $\varphi^{\text{ToP}}_{n}$ is a good initial guess for numerically finding the $\gamma=1$ ToP$(n+1)$ root to $\mathcal{B}_{n,2n}^{\text{AP}}$. In this manner, we obtain the necessary zero-order solutions for continuation over all $\gamma$.

\begin{figure}
\includegraphics[width=\columnwidth]{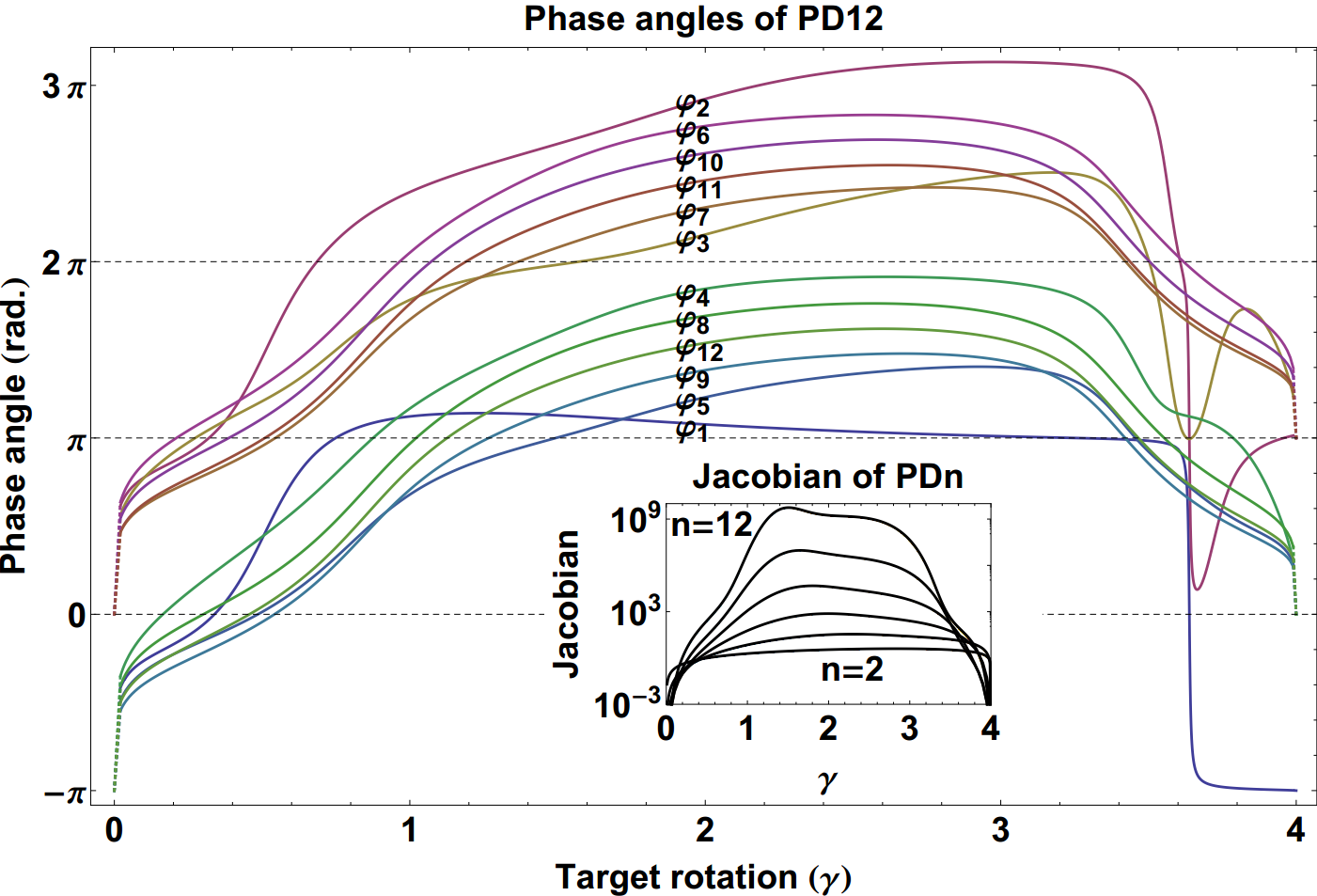}
\caption{\label{PD12plot} Phase angles $\varphi_k$ as a function of target rotation $\gamma\in(0,4)$ for a length $24$ order $12$ palindromic PD$12$ sequence. This example demonstrates the method of analytic continuation from an exact solution $\left.\vec{\varphi}^{\text{PD}}_{12,6}\right|_{\gamma=2}=\frac{12\pi}{13}(1,-1,2,-2,3,-3,4,-4,5,-5,6,-6)$, a procedure valid wherever the Jacobian of a sequence is non-zero, also plotted for $\vec{\varphi}^{\text{PD}}_{n,n/2}$ (inset). The dotted lines represent the expected continuation to the non-analytic points $\gamma=0,4$.}
\end{figure}

\subsubsection{Analytic continuation of PD$n$}
\label{Analytic continuation of PD$n$}

We would like to find for PD$n$ a set of initial points $(\vec{\varphi}_0,\gamma_0$) that are suitable for analytic continuation. These are provided by inspecting the PD$2,4$ closed from solutions at $\gamma=2$ which are of the form $\vec{\varphi}^{\text{PD}}_{n}=\frac{2 \pi}{n+1}(1,-1,2,-2,...,n/2,-n/2)$. We prove that these are arbitrary $n$ solutions by noting the sequence
\begin{equation}
\label{vitanov_sequences}
V_n=\frac{2 \pi}{n+1}(0,1,-1,2,-2,...,n/2,-n/2),
\end{equation}
without palindrome symmetry applied is in fact the length $L=n+1$ Class-B \cite{Levitt1986} sequence reported by Vitanov \cite{Vitanov2011} that is optimal in the sense that for all odd $j<L$, $\Phi_L^j(\vec\varphi)=0$ is satisfied. Thus by applying palindrome symmetry and dropping the zeroth pulses, a length $L=2n$, $\gamma=2$ Class-A sequence is obtained. We conjecture that $\vec{\varphi}^{\text{PD}}_{n}$ is contained in the more general class of $\gamma=2$ solutions
\begin{equation}
\label{PDn_Initial_values}
\left.\vec{\varphi}^{\text{PD}}_{n,m}\right|_{\gamma=2}=\frac{2 m\pi}{n+1}(1,-1,2,-2,...,n/2,-n/2),
\end{equation}
where $(n+1,m)$ are coprime.

Arbitrary angle solutions to PD$n$, such as in Fig. \ref{PD12plot}, may be obtained by continuation from the $\left.\vec{\varphi}^{\text{PD}}_{n,m}\right|_{\gamma=2}$. Unlike the ToP$n$ sequences, the Jacobian plotted in Fig. \ref{PD12plot} (inset) for instances of $\vec{\varphi}^{\text{PD}}_{n,n/2}$ is non-zero over the entire range of $\gamma$ up to $n=12$.

\subsubsection{Brute-force numerical search}
\label{Brute-force numerical search}

Another solution method is a direct numerical search for all possible real solutions to Eq.~\eqref{phase_sum=ffunction} for a given $\gamma$. This is by far the least efficient approach, but allows one to obtain other classes of sequences that are not contained in $\vec{\varphi}^{\text{ToP}}_{n}$ and $\vec{\varphi}^{\text{PD}}_{n,m}$. Searching up to $n=8$ suggests that, after palindrome or antipalindrome symmetry has been applied, for every value of $\gamma$, $2^{\lceil n/2\rceil}$ distinct real sequences exist. These sequences may also be analytically continued, and we provide in Table~\ref{numerical_varphi_table} their phase angles at $\gamma=\{1,\frac{1}{2},\frac{1}{4}\}$ with which the interested reader may use to do so. In the table, we also provide a measure of the leading order error of each sequence $\kappa$, where $\kappa$ is defined such that, if the leading error from Eq.~\eqref{leading_error} is considered a small rotation, its amplitude is $(\kappa\epsilon\theta_0/2)^{n+1}$.

\section{Further extensions}
\label{section_Further extensions}
\begin{figure}
\includegraphics[width=\columnwidth]{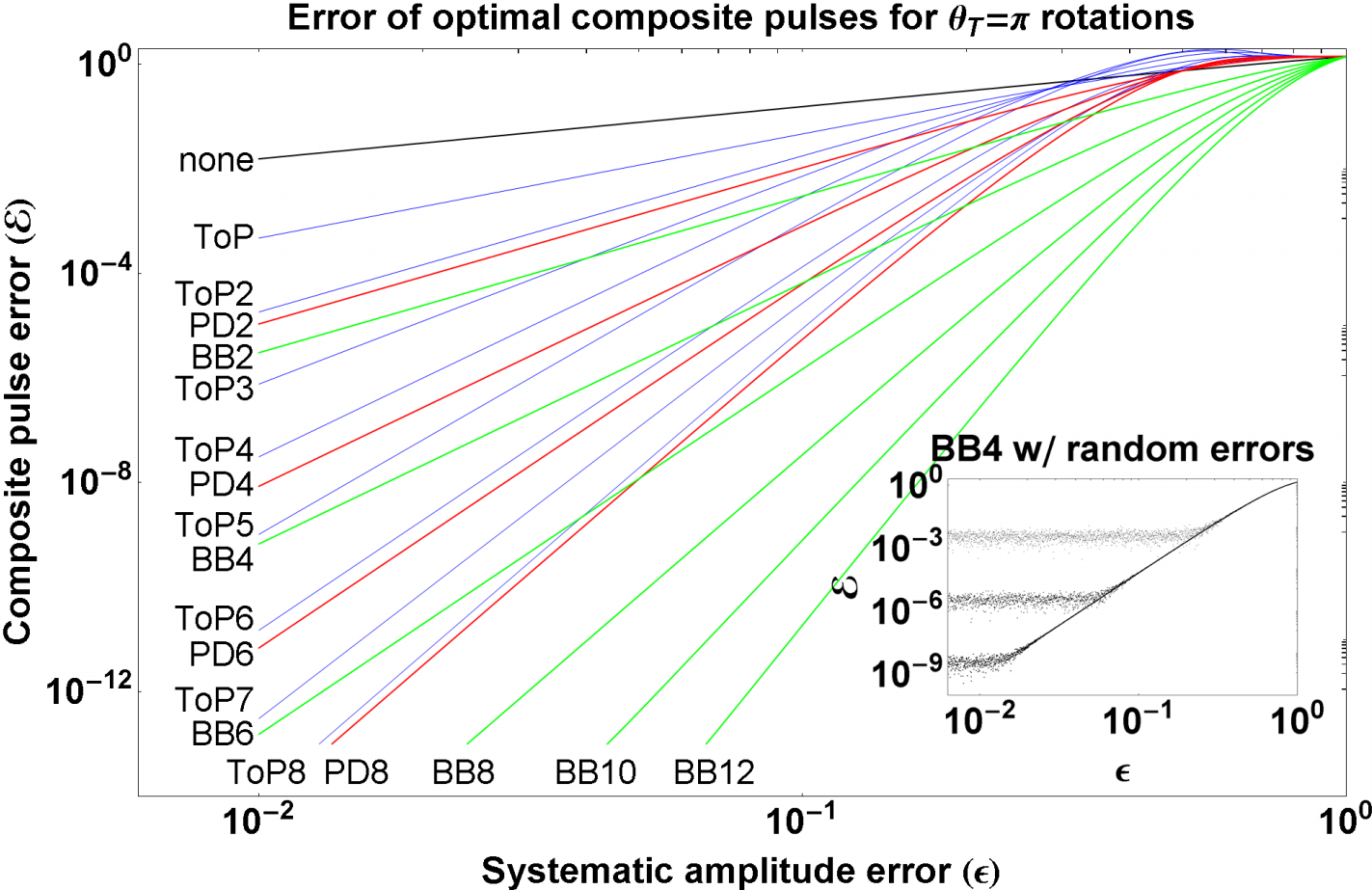}
\caption{\label{tracerrror} The trace distance of order $n$ sequences $U_T$ from an ideal $\pi$-rotation, $\mathcal{E}=\left\lVert U_T - R_0[\pi]\right\rVert$, as a function of error $\epsilon$. Included are $\theta_0=2\pi$ passband sequences ToP$n$ (blue) and PD$n$ (red), as well as the broadband $\theta_0=\pi$ BB$n$ (green) obtained by `toggling' PD$n$ phase angles. Observe that the asymptotic gradient is $\mathcal{O}(\epsilon^{n})$, and that the number of $2\pi$ imperfect rotations $=n\theta_0/\pi$ given $n$ is dramatically shorter than prior work by Brown et al. \cite{Brown2004}. The inset demonstrates the effects of experimental imprecisions by plotting $\mathcal{E}$ for a population of BB$4$ sequences subject to combined Gaussian distributed amplitude and phase errors with zero mean and standard deviation $\sigma=10^{-3,-6,-9}$ from top.}
\end{figure}

Some generalizations of the above results are possible. First, the sequence duration $L\theta_0/2\pi$ may be halved by considering $\theta_0=\pi$ sequences, with a significant decrease in the trace distance $\cal{E}$ from an ideal rotation as demonstrated in  Fig.~\ref{tracerrror}. This leads to similar Eqs.~(\ref{bb_constraint_2pi}-\ref{leading_error}), but with replacements ${\Phi_L^j(\vec\varphi)\rightarrow\Phi_L^{L-j}(\vec\psi)}$ and ${e^{ix\gamma}\rightarrow (-i)^Le^{ix\gamma}}$. The $x^0$ term of the counterpart to Eq.~\eqref{bb_constraint_2pi} implies that $L\in 4\mathbf{Z}$.

Second, while all $2\pi$-pulse sequences are passband \cite{Wimperis1994}, broadband $\pi$-pulse sequences BB$n$ with phase angles $\vec\psi$ can be obtained from palindrome sequences PD$n$ with phase angles $\vec\varphi$ through the ``toggling" transformation ${\psi_k=-\sum_{h=1}^{k-1}(-1)^h\varphi_h+\sum_{h=k+1}^L(-1)^h\varphi_h}$. Narrowband $\pi$-pulse sequences are obtained simply by using the original phase angles $\vec\varphi$.

Third, nonlinear systematic amplitude error suppression is also possible. Say the erroneous rotations making up sequence $\mathcal{S}$ were instead $M_\varphi[\theta_0]=R_\varphi[\theta_0+\epsilon E(\theta_0)]$ for any function $E$ with $E(\theta_0)\neq0$. Then, the normalization $x\equiv\frac12\epsilon E(\theta_0)$ and $\gamma\equiv E(\theta_T)/E(\theta_0)$ preserves Eqs.~(\ref{bb_constraint2a}-\ref{leading_error}). Therefore, our pulse sequences correct nonlinear errors to $\mathcal{O}(\epsilon^n)$, assuming the ratio $E(\theta_T)/E(\theta_0)$ is known.

Real applications involve additional systematic and random errors, assumed to be small, in both system and control. Examples of such include phase errors in $\varphi$ as well as off-resonance errors \cite{Merrill2012}. These contribute to the primitive pulse $M_{\varphi}[\theta]$ a small effective rotation $R_{\vec{n}}[\mathcal{O}(\delta)],\, |\delta|\ll 1$, about some arbitrary axis $\vec{n}$. Note that interchanging the order $R_{\vec{n}}[\mathcal{O}(\delta)]\text{ }M_{\varphi}[\theta]=M_{\varphi}[\theta]\text{ }R_{\vec{m}}[\mathcal{O}(\delta)]$ only changes the rotation axis. Including uncorrected errors, our sequences become ${\mathcal{S}'=\prod_{j=1}^L R_{\vec{n}_{j}}[\mathcal{O}(\delta_{j})]\text{ }M_{\varphi_j}[\theta_0]=\left(\prod_{j=1}^L R_{\vec{m}_{j}}[\mathcal{O}(\delta_j)]\right)\mathcal{S}}$. Hence, the amplitude error suppression property of a pulse sequence is preserved, whilst amplifying other sources of error to $L \mathcal{O}(\delta)$. If $\delta$ is purely random with variance $\sigma^{2}$, this imposes a practical limit on the order $n$ of error suppression $\sqrt{L} \mathcal{O}(\sigma)\sim\mathcal{O}(\epsilon^{n})$, beyond which no decrease in pulse error is obtainable. This further implies that our sequences can be implemented even with significant experimental imprecision, which correspond to at most linearly accumulating errors that simply level off at small $\epsilon$, as illustrated in Fig.~\ref{tracerrror} (inset). Thus, in general, efficient sequences amplify uncorrected sources of error by at most $L=\operatorname{poly}(n)$.

Since amplitude error suppression to order $n$ is only effective when other errors are small $|\delta| \sim |\epsilon|^{n}$, the simultaneous suppression of multiple sources of systematic errors is highly desirable. This can often be achieved by sequence concatenation \cite{Merrill2012,Bando2013}. For example, let AM$n=\prod_{k=1}^LM_{\varphi_k}[\theta_k]$ denote any sequence that suppresses systematic amplitude errors to order $n$. Noting that a $\theta=2\pi$ rotation is already robust to off-resonance errors \cite{Bando2013}, replacing each $\theta\neq2\pi$ pulse in AM$n$ by a CORPSE sequence \cite{Cummins2003} results in a concatenated sequence suppressing off-resonance errors to first order and amplitude errors to order $n$. This follows from the observation that the CORPSE sequence approximating $R_{\varphi_T}[\theta_T]$ yields the rotation $M_{\varphi_T}[\theta_T]\cdot R_{\vec n}[\mathcal{O}(\delta^2+\delta\epsilon)]$ for some axis $\vec n$, where $\delta$ is the order of the off-resonance error \cite{Bando2013}. This property is special to CORPSE, because the decomposition into $M_{\varphi_T}[\theta_T]$ times a small error will not generally occur.

\section{Comparison with prior art}
\label{section_Comparison with prior art}
There are very few sequences in the literature that allow for corrected arbitrary angle rotations (criterion (4) from our introduction). The classic examples SCROFULOUS \cite{Cummins2003} and the PB sequences \cite{Wimperis1994} are only correct to $n=1,2$ respectively. The work of Brown et. al. \cite{Brown2004} report sequences SK$n$ for arbitrary $n$, but with a length scaling of $L=\mathcal{O}(n^3)$. Thus, for any given $n$, the class of optimal $L=2n$ sequences PD$n$, AP$n$ correct to same order as SK$n$, but with significantly fewer pulses, as seen by comparing their trace errors in Fig. \ref{tracerrror} with those of Brown \cite{Brown2004}.

Therefore, for a more comprehensive comparison with prior art, we limit ourselves to inverting sequences so that $\theta_0=\theta_T=\pi$ and the initial state is always $\ket{0}$. This corresponds to toggled sequences for $\gamma=1$ in Table.~\ref{numerical_varphi_table}, and corresponds to relaxing criteria (3) and (4) from our introduction. For consistency of notion, the pulse length $L$ will now include the zeroth pulse. 

First, we note that the leading arbitrary accuracy $\theta_T=\pi$ Class A sequences in the literature are derived by recursive nesting \cite{Husain2013}: Given some order $n_b$ base sequence comprised of $L_b$ $\pi$ pulses, one performs a nesting procedure \cite{Jones2013} $k$ times to obtain an order $n=(n_b+1)^k-1$ sequence with length $L=L_b^k$, corresponding to a length scaling of $L=(n+1)^{\log{L_b}/\log(n_b+1)}$. Wimperis' BB$_1$ sequence \cite{Wimperis1994} with $L_b=5, n_b=2$ is often used as the base sequence, resulting in the $F_r$ class of length $L=5^r$, order $n=3^r-1$ sequences \cite{Husain2013}, with a asymptotic scaling of $L=\mathcal{O}(n^{1.47})$, slightly worse than that of the $L=2n+1$ of BB$n$ sequences reported here. Furthermore, our BB$n$ sequences at $\gamma=1$ can themselves be nested after rearranging in the toggled $\theta_0=2\pi$ frame such that the zeroth pulse is in the middle,
\begin{align}
\mathcal{S}_{2\pi}\cdot M_{0}=&M_{\varphi_1}M_{\varphi_2}\cdots M_{\varphi_2}M_{\varphi_1}M_{0}=R_0[2\pi]+\mathcal{O}(\epsilon^{n+1})\\\nonumber
=& M_{\varphi_n}\cdots M_{\varphi_1}M_{0}M_{\varphi_1}\cdots M_{\varphi_n}+\mathcal{O}(\epsilon^{n+1}),
\end{align}
For example, if BB$12$ is used as the base sequence, $L=\mathcal{O}(n^{1.25})$ is achieved. In the limit of large $L_b=2 n_b+1$, $L$ asymptotically approaches $\mathcal{O}(n)$.

\begin{figure}
\includegraphics[width=\columnwidth]{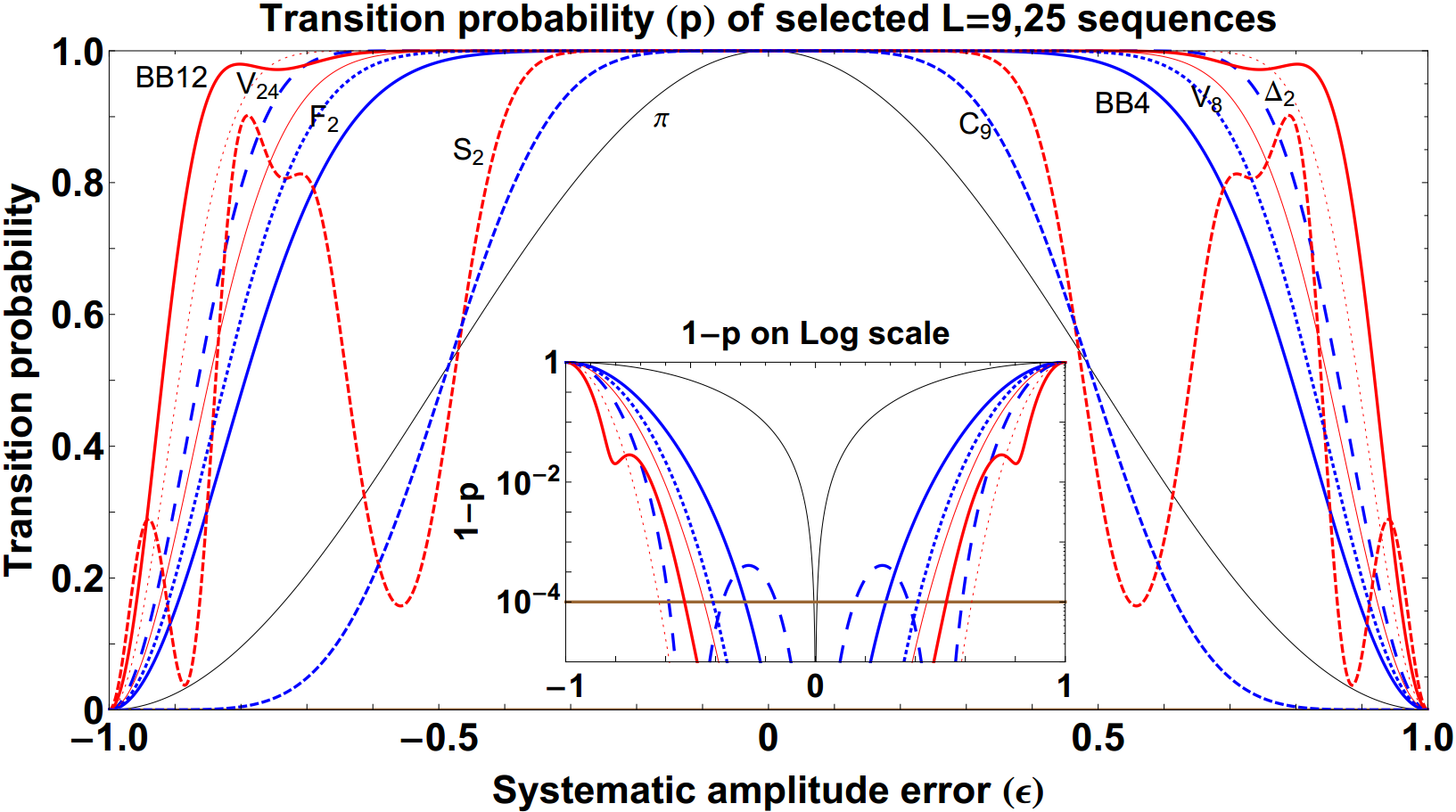}
\caption{\label{comparisonplot} Comparison of the transition probability $p=|\left\langle1\right|U_T\left|0\right\rangle|^2$ with amplitude error $\epsilon$ between a variety of $L=9$ (blue) and $L=25$ (red) pulse sequences implementing $\pi$ rotations. Solid(dashed) lines represent Class A(B) sequences. Included are the BB$4,12$ of this work, $V_{8,24}$ (Eq. \ref{vitanov_sequences}), $\Delta_2$ \cite{Shaka1984}, $C_9$ \cite{Cho1984}, F$_2$ \cite{Husain2013} and $S_2$\cite{Tycko1985}. In the inset, $1-p$ is plotted on a logarithmic scale.}
\end{figure}

Unlike Class A sequences, Class B sequences are less interesting from the context of quantum computing as they require specific initial states. The trace error metric is inapplicable, but one can nevertheless plot the transition probability $|\bra{1} U_T \ket{0}|^2$ as a function of $\epsilon$ for $\theta_T=\pi$ rotations acting on the ground state as in Fig. \ref{comparisonplot}. As our highest order closed-form sequence BB$4$ specialized to $\theta_T=\pi$ has $L=9$, we perform a comparison with $L=9$ inverting sequences in the literature: $V_8$ by Vitanov \cite{Vitanov2011}, $\Delta_2=\frac{\pi}{6}(0, 3, 0, 4, 7, 4, 0, 3, 0)$ by Shaka et. al \cite{Shaka1984} and $C_9=\frac{\pi}{12}(0, 1, 12 , 11 , 18 , 11 , 12 , 1, 0)$ by Cho et. al \cite{Cho1984}. Other than ours, there do not appear to be any $L=9$ Class A sequences. We also include a few $L=25$ sequences: the Class A BB$12$ and F$_2$ \cite{Wimperis1994,Husain2013}, and the inverting sequences $V_{24}$ and $S_2=\frac{\pi}{3}(0, 0, 2, 1, 2, 0, 0, 2, 1, 2, 2, 2, 4, 3, 4, 1, 1, 3, 2, 3, 2, 2, 4,
  3, 4)$ \cite{Tycko1985}. In this notation, $\theta_0=\pi$ and, for example, $C_9$ is implemented by $M_{0}M_{\frac{\pi}{12}}M_{\pi}\cdots$. We see that even against the specialized inverting sequences of similar length, BB$n$ compares favorably. At the $10^{-4}$ quantum error threshold, the width of BB$n$ about $\epsilon=0$ is only outperformed by the optimal inverting sequences $V_n$ \cite{Vitanov2011}.
  
\section{Conclusion}
\label{section_Conclusion}
The study of pulse sequences is a broad discipline, covering many different pulse shapes, error models, and computational methods, and we have dealt here with only a small of this breadth. However, our algebraic approach to amplitude errors provides a major characterization of this important and ubiquitous case. The constraints in Eq.~\eqref{constraints} are both necessary and sufficient for any $2\pi$-pulse sequence of length $L$ correcting to order $n$. Using these constraints we were able to find sequences in closed form beyond any order previously known analytically. From our $2\pi$-pulse passband sequences, we also demonstrated that it is simple to obtain $\pi$-pulse broadband and $\pi$-pulse narrowband sequences of the same optimal length.

In the introduction, we proposed four criteria for a pulse sequence to be useful to quantum computation: (1) It exists at all orders of correction $n$. (2) It has efficient length $L=\mathcal{O}(\operatorname{poly}(n))$. (3) It can operate as intended on any initial state. (4) It exists for all target rotations. We can evaluate our results with respect to these criteria. We proved that our ToP$n$ and PD$n$ sequences satisfy (1), (2), and (3) and provided evidence that (4) holds for them as well by studying the Jacobian of the phase sums and proving that ToP$n$ solutions do indeed exist for a range of target rotations around the identity. At the same time, our numerical results indicate that conditions (2), (3), and (4) hold up to order $n=12$ with $L=2n$. This is a strong start to a complete proof that all four criteria are satisfiable for amplitude error correcting sequences. In fact, even a weaker proof that criteria (1), (2), and (3) hold at target angles $\pi$ and $\pi/4$ would be valuable, as then pulse sequences for the Hadamard and T-gate, together sufficient for universal single-qubit computation \cite{Nielsen2004}, could be implemented.

A very natural extension of our approach is to find constraints for sequences correcting other systematic control errors, for instance, arbitrary system drifts and off-resonance errors. As an example, Uhrig's \cite{Uhrig2007} algebraic approach to dephasing errors might also be amenable to generalization. An anticipated problem in these other cases might be the non-polynomial nature of such constraint equations, meaning more difficult analytic and numeric methods might be needed to find solutions.

This work was supported by the NSF CUA, NSF iQuISE IGERT, IARPA QCS ORAQL, and NSF CCF-RQCC projects.

\appendix
\setcounter{equation}{0}
\renewcommand{\theequation}{A.\arabic{equation}}
\section*{Appendix: Gr\"{o}bner Bases and Buchberger's Algorithm}
As the method of Gr\"{o}bner bases is likely not well-known to expected readers, we believe that a demonstration of deriving the bases for AP$1$ and PD$2$ would be instructive. Generalization to AP$3$, PD$4$ is straightforward, but more computationally intensive. To begin, we use a definition of the Gr\"{o}bner basis that facilitates its computation \cite{Buchberger2006}:
\begin{definition}
\label{GBdefinition}
Given a \emph{term order}, a set $G$ is a Gr\"{o}bner basis if and only if for all $f,g\in G$ the S-polynomial $\mathrm{SPOL}(f,g)$ by repeated reduction with respect to $G$ can be brought to zero.
\end{definition}
This definition follows from Buchberger's theorem and leads to his famous algorithm for computing a Gr\"{o}bner basis of $\mathcal{F}$ \cite{Buchberger2006}:
\begin{algorithmic}
\Function{Buchberger's algorithm}{$\mathcal{F}$}
\State $\mathcal{G} \gets \mathcal{F}$\Comment{System of multivariate polynomials}
\State $C\gets\mathcal{G}\times \mathcal{G}$
\While{$C\neq \emptyset$}
    \State $p\gets (a,b)\in C$\Comment{Chosen arbitrarily}
    \State $C\gets C\backslash\{p\}$
    \State $h=\mathrm{RED}(\mathrm{SPOL}(a,b),\mathcal{G})$\Comment{Defined below}
    \If {$h\neq0$}
        \State $C\gets C\cup(\mathcal{G}\times\{h\})$
        \State $\mathcal{G}\gets \mathcal{G}\cup\{h\}$
    \EndIf
\EndWhile
\State \Return $\mathcal{G}$\Comment{Gr\"{o}bner basis of $\mathcal{F}$}
\EndFunction
\end{algorithmic}
where
\begin{align}\nonumber
\mathrm{RED}(a,\mathcal{G})=&\text{Remainder of }a \text{ upon division by }\mathcal{G} \text{(reduction)},\\\nonumber
\mathrm{SPOL}(a,b)=&lcm\left(LPP(a),LPP(b)\right)\left(\frac{a}{LM(a)}-\frac{b}{LM(b)}\right),\\\nonumber
lcm(a,b)=&\text{least common multiple of }a,b,\\\nonumber
LM(a)=&\text{leading monomial of }a\text{ w.r.t. some term order},\\\nonumber
LPP(a)=&LM(a)\text{ with coefficients dropped}.\nonumber
\end{align}

In what follows, we apply the Weierstrass substitution $\tan(\varphi_k/2)=t_k$ to $\mathcal{B}_{n,L}$, rearrange to obtain a polynomial system $\mathcal{W}_{n,L}$, and use the lexicographic monomial ordering $t_1 \prec_{lex} t_2 \prec_{lex}\cdots\prec_{lex} t_n$ \cite{Cox2007} in computing a Gr\"{o}bner basis.

\subsection{Example: AP$1$ from $\mathcal{B}_{1,2}^{\text{AP}}$}

$\mathcal{B}_{1,2}^{\text{AP}}\Rightarrow\mathcal{W}_{1,2}^{\text{AP}}=\{t_1^2\left(1+\frac{\gamma}{2}\right)-(1-\frac{\gamma}{2})\}$. This example is trivial as $\mathcal{W}_{1,2}^{\text{AP}}$ is automatically a Gr\"{o}bner basis $\mathcal{G}$ following from Definition \ref{GBdefinition} as the S-polynomial of an arbitrary polynomial with itself is $0$. As $\mathcal{G}$ generates $\mathcal{W}_{1,2}^{\text{AP}}$, they share the same simultaneous roots. Solving $\mathcal{G}$, we obtain $t_1=\pm\sqrt{\frac{2-\gamma}{2+\gamma}}$. Solving for $\varphi_1=\cos^{-1}{\left(\frac{\gamma}{2}\right)}$, we see that AP$1$ is the sequence SK$1$ \cite{Brown2004}.

\subsection{Example: PD$2$ from $\mathcal{B}_{2,4}^{\text{PD}}$}

$\mathcal{B}_{2,4}^{\text{PD}}\Rightarrow\mathcal{W}_{2,4}^{\text{PD}}=\{t_1^2 t_2^2(\gamma -4)+t_1^2\gamma+t_2^2\gamma+(4+\gamma),t_1^2 t_2+t_1 t_2^2+t_1+t_2\}$. In rearranging, we have introduced the complex roots $1+t_k^2=0$ which we shall have to remove later. One could solve $\mathcal{W}_{2,4}^{\text{PD}}$ by inspection, but we apply Buchberger's algorithm to demonstrate the algorithmic manner in which solutions may be derived. We perform the first iteration in detail and only state the computed basis element $h_{i,j}$ of succeeding iterations for brevity:
\begin{align}
\label{PD2 example}
&\text{Input: }\mathcal{W}_{2,4}^{\text{PD}}.\\\nonumber
&\mathcal{G}=\mathcal{W}_{2,4}^{\text{PD}},\, C=\{(g_1,g_2)\},\\\nonumber
&\quad \text{Take the pair }g_1,g_2.\\\nonumber
&\quad LM(g_1)=t_2^2 t_1^2 (\gamma -4)=(\gamma -4)LPP(g_1),\\\nonumber
&\quad LM(g_2)=t_2^2 t_1 = LPP(g_2),\\\nonumber
&\quad LCM(LPP(g_1),LPP(g_2))=t_2^2 t_1^2,\\\nonumber
&\quad SPOL(g_1,g_2)=\frac{\gamma  t_ 2^2}{\gamma -4}-t_1^3 t_ 2-t_1 t_ 2+\frac{4 t_1^2}{\gamma -4}+\frac{\gamma +4}{\gamma -4},\\\nonumber
&\quad h_{1,2}=RED(SPOL(g_1,g_2),\mathcal{G})=SPOL(g_1,g_2),\\\nonumber
&\mathcal{G} = \mathcal{G} \cup \{h_{1,2}\},\,C=\{(g_1,g_3),(g_2,g_3)\},\\\nonumber
&\cdots\\\nonumber
&\text{Output: }\mathcal{G}=\mathcal{W}_{2,4}^{\text{PD}}\cup\{h_{1,2},h_{2,3},h_{1,4},h_{4,5},h_{1,6}\},\\\nonumber
&h_{2,3}=\frac{(\gamma -4) t_ 2 t_ 1^4+2 (\gamma -2) t_ 2 t_ 1^2 +\gamma  t_ 2-4 t_1^3-4 t_ 1}{\gamma},\\\nonumber
&h_{1,4}=\frac{4 t_ 1^3 t_ 2+4 t_ 1 t_ 2+\gamma  t_ 1^4+2 (\gamma +2) t_ 1^2+\gamma +4}{\gamma-4},\\\nonumber
&h_{4,5}=\frac{4 \gamma   t_ 2 t_ 1^2 +  4 \gamma  t_ 2 - (\gamma - 4) \gamma  t_ 1^5 -  2 (\gamma - 2) \gamma  t_ 1^3 - \gamma ^2 t_ 1}{4(4-\gamma)},\\\nonumber
&h_{1,6}=\frac{(\gamma -4) t_ 1^6+(3 \gamma -4) t_ 1^4+(3 \gamma +4) t_ 1^2+\gamma +4}{4},\\\nonumber
&h_{i,j}=0\quad\forall i,j = 1,...,|\mathcal{G}|.\\\nonumber
\end{align}
Note that the last term $h_{1,6}$ is univariate in $t_1$, as expected from a regular chain. We have chosen pairs from $C$ so as to minimize the output, but any arbitrary choice will eventually terminate. However $\mathcal{G}$ from Eq.~\ref{PD2 example} still contains more elements than is necessary to generate $\mathcal{W}_{2,4}^{\text{PD}}$. We can deterministically compute from $\mathcal{G}$ a unique minimal, or \emph{reduced}, Gr\"{o}bner basis $\mathcal{G}_R$ up to constant factors by repeating $\mathcal{G}\leftarrow(\mathcal{G}-\{g\})\cup\{RED(g,\mathcal{G}-\{g\})\}\ \forall g\in\mathcal{G}$ until the process converges \cite{Cox2007}:
\begin{align}
\mathcal{G}_R=&\left\{\left(t_1^2+1\right)^2 \left(4+\gamma-(4-\gamma) t_1^2\right)\right.,\\\nonumber
&\left(t_1^2+1\right) \left((\gamma -4) t_1^3+\gamma  t_1-4 t_2\right),\\\nonumber
&\left.4+\gamma +(\gamma -4) t_1^4+2 (\gamma -2) t_1^2+4 t_2^2)\right\}.
\end{align}

The non-physical zeroes $\mathcal{M}=\{1+t_1^2,1+t_2^2\}$ that were introduced earlier are now apparent and can be removed. We can deterministically compute from $\mathcal{G}_R$ another Gr\"{o}bner basis $\mathcal{G}_Q$  with the same zeros sans $\mathcal{M}$ by repeatedly computing the Gr\"{o}bner basis of the ideal quotient $\mathcal{G_Q}\rightarrow\langle \mathcal{G_Q}\rangle:\langle\mathcal{M}\rangle$ until convergence, or \emph{saturation} \cite{Cox2007}. Finally, we obtain the simple triangular system
\begin{align}
\mathcal{G}_Q=\{4+\gamma -(4-\gamma) t_1^2,t_1+t_2\}.
\end{align}
Solving for $\varphi_1=\cos^{-1}{\left(\frac{\gamma}{4}\right)},\,\varphi_2=-\varphi_1$, we see that PD$2$ is the sequence PB$_1$ \cite{Wimperis1994}.

\bibliography{PulseSequencesMin}

\begin{thebibliography}{26}
\expandafter\ifx\csname natexlab\endcsname\relax\def\natexlab#1{#1}\fi
\expandafter\ifx\csname bibnamefont\endcsname\relax
  \def\bibnamefont#1{#1}\fi
\expandafter\ifx\csname bibfnamefont\endcsname\relax
  \def\bibfnamefont#1{#1}\fi
\expandafter\ifx\csname citenamefont\endcsname\relax
  \def\citenamefont#1{#1}\fi
\expandafter\ifx\csname url\endcsname\relax
  \def\url#1{\texttt{#1}}\fi
\expandafter\ifx\csname urlprefix\endcsname\relax\def\urlprefix{URL }\fi
\providecommand{\bibinfo}[2]{#2}
\providecommand{\eprint}[2][]{\url{#2}}

\bibitem[{\citenamefont{Nielsen and Chuang}(2004)}]{Nielsen2004}
\bibinfo{author}{\bibfnamefont{M.~A.} \bibnamefont{Nielsen}} \bibnamefont{and}
  \bibinfo{author}{\bibfnamefont{I.~L.} \bibnamefont{Chuang}},
  \emph{\bibinfo{title}{{Quantum Computation and Quantum Information}}}
  (\bibinfo{publisher}{Cambridge University Press}, \bibinfo{year}{2004}),
  \bibinfo{edition}{1st} ed.

\bibitem[{\citenamefont{Merrill and Brown}(2012)}]{Merrill2012}
\bibinfo{author}{\bibfnamefont{J.~T.} \bibnamefont{Merrill}} \bibnamefont{and}
  \bibinfo{author}{\bibfnamefont{K.~R.} \bibnamefont{Brown}},
  \bibinfo{type}{Tech. Rep.} \bibinfo{number}{arXiv:1203.6392}
  (\bibinfo{year}{2012}).

\bibitem[{\citenamefont{Tycko et~al.}(1985)\citenamefont{Tycko, Pines, and
  Guckenheimer}}]{Tycko1985}
\bibinfo{author}{\bibfnamefont{R.}~\bibnamefont{Tycko}},
  \bibinfo{author}{\bibfnamefont{A.}~\bibnamefont{Pines}}, \bibnamefont{and}
  \bibinfo{author}{\bibfnamefont{J.}~\bibnamefont{Guckenheimer}},
  \bibinfo{journal}{The Journal of Chemical Physics}
  \textbf{\bibinfo{volume}{83}}, \bibinfo{pages}{2775} (\bibinfo{year}{1985}).

\bibitem[{\citenamefont{Levitt}(1986)}]{Levitt1986}
\bibinfo{author}{\bibfnamefont{M.~H.} \bibnamefont{Levitt}},
  \bibinfo{journal}{Progress in Nuclear Magnetic Resonance Spectroscopy}
  \textbf{\bibinfo{volume}{18}}, \bibinfo{pages}{61 } (\bibinfo{year}{1986}).

\bibitem[{\citenamefont{Brown et~al.}(2004)\citenamefont{Brown, Harrow, and
  Chuang}}]{Brown2004}
\bibinfo{author}{\bibfnamefont{K.~R.} \bibnamefont{Brown}},
  \bibinfo{author}{\bibfnamefont{A.~W.} \bibnamefont{Harrow}},
  \bibnamefont{and} \bibinfo{author}{\bibfnamefont{I.~L.}
  \bibnamefont{Chuang}}, \bibinfo{journal}{Phys. Rev. A}
  \textbf{\bibinfo{volume}{70}}, \bibinfo{pages}{052318}
  (\bibinfo{year}{2004}).

\bibitem[{\citenamefont{Ichikawa et~al.}(2012)\citenamefont{Ichikawa, Bando,
  Kondo, and Nakahara}}]{Ichikawa2012}
\bibinfo{author}{\bibfnamefont{T.}~\bibnamefont{Ichikawa}},
  \bibinfo{author}{\bibfnamefont{M.}~\bibnamefont{Bando}},
  \bibinfo{author}{\bibfnamefont{Y.}~\bibnamefont{Kondo}}, \bibnamefont{and}
  \bibinfo{author}{\bibfnamefont{M.}~\bibnamefont{Nakahara}},
  \bibinfo{journal}{Philosophical Transactions of the Royal Society A:
  Mathematical,Physical and Engineering Sciences}
  \textbf{\bibinfo{volume}{370}}, \bibinfo{pages}{4671} (\bibinfo{year}{2012}).

\bibitem[{\citenamefont{Jones}(2013)}]{Jones2013}
\bibinfo{author}{\bibfnamefont{J.~A.} \bibnamefont{Jones}},
  \bibinfo{journal}{Physics Letters A} \textbf{\bibinfo{volume}{377}},
  \bibinfo{pages}{2860 } (\bibinfo{year}{2013}).

\bibitem[{\citenamefont{Husain et~al.}(2013)\citenamefont{Husain, Kawamura, and
  Jones}}]{Husain2013}
\bibinfo{author}{\bibfnamefont{S.}~\bibnamefont{Husain}},
  \bibinfo{author}{\bibfnamefont{M.}~\bibnamefont{Kawamura}}, \bibnamefont{and}
  \bibinfo{author}{\bibfnamefont{J.~A.} \bibnamefont{Jones}},
  \bibinfo{journal}{Journal of Magnetic Resonance}
  \textbf{\bibinfo{volume}{230}}, \bibinfo{pages}{145 } (\bibinfo{year}{2013}).

\bibitem[{\citenamefont{Cummins et~al.}(2003)\citenamefont{Cummins, Llewellyn,
  and Jones}}]{Cummins2003}
\bibinfo{author}{\bibfnamefont{H.~K.} \bibnamefont{Cummins}},
  \bibinfo{author}{\bibfnamefont{G.}~\bibnamefont{Llewellyn}},
  \bibnamefont{and} \bibinfo{author}{\bibfnamefont{J.~A.} \bibnamefont{Jones}},
  \bibinfo{journal}{Phys. Rev. A} \textbf{\bibinfo{volume}{67}},
  \bibinfo{pages}{042308} (\bibinfo{year}{2003}).

\bibitem[{\citenamefont{Wimperis}(1994)}]{Wimperis1994}
\bibinfo{author}{\bibfnamefont{S.}~\bibnamefont{Wimperis}},
  \bibinfo{journal}{J. Mag. Reson., Ser A} \textbf{\bibinfo{volume}{109}},
  \bibinfo{pages}{221 } (\bibinfo{year}{1994}).

\bibitem[{\citenamefont{Viola et~al.}(1999)\citenamefont{Viola, Knill, and
  Lloyd}}]{Viola1999}
\bibinfo{author}{\bibfnamefont{L.}~\bibnamefont{Viola}},
  \bibinfo{author}{\bibfnamefont{E.}~\bibnamefont{Knill}}, \bibnamefont{and}
  \bibinfo{author}{\bibfnamefont{S.}~\bibnamefont{Lloyd}},
  \bibinfo{journal}{Phys. Rev. Lett.} \textbf{\bibinfo{volume}{82}},
  \bibinfo{pages}{2417} (\bibinfo{year}{1999}).

\bibitem[{\citenamefont{Khodjasteh et~al.}({2010})\citenamefont{Khodjasteh,
  Lidar, and Viola}}]{Khodjasteh2010}
\bibinfo{author}{\bibfnamefont{K.}~\bibnamefont{Khodjasteh}},
  \bibinfo{author}{\bibfnamefont{D.~A.} \bibnamefont{Lidar}}, \bibnamefont{and}
  \bibinfo{author}{\bibfnamefont{L.}~\bibnamefont{Viola}},
  \bibinfo{journal}{Phys. Rev. Lett.} \textbf{\bibinfo{volume}{{104}}},
  \bibinfo{pages}{{090501}} (\bibinfo{year}{{2010}}).

\bibitem[{\citenamefont{Souza}(2012)}]{Souza2012}
\bibinfo{author}{\bibfnamefont{A.}~\bibnamefont{Souza}},
  \bibinfo{journal}{Philosophical Transactions of the Royal Society A:
  Mathematical,Physical and Engineering Sciences}
  \textbf{\bibinfo{volume}{370}}, \bibinfo{pages}{4748} (\bibinfo{year}{2012}).

\bibitem[{\citenamefont{Uhrig}({2007})}]{Uhrig2007}
\bibinfo{author}{\bibfnamefont{G.~S.} \bibnamefont{Uhrig}},
  \bibinfo{journal}{Phys. Rev. Lett.} \textbf{\bibinfo{volume}{{98}}},
  \bibinfo{pages}{{100504}} (\bibinfo{year}{{2007}}).

\bibitem[{\citenamefont{Vitanov}(2011)}]{Vitanov2011}
\bibinfo{author}{\bibfnamefont{N.~V.} \bibnamefont{Vitanov}},
  \bibinfo{journal}{Phys. Rev. A} \textbf{\bibinfo{volume}{84}},
  \bibinfo{pages}{065404} (\bibinfo{year}{2011}).

\bibitem[{\citenamefont{Gelfand et~al.}(1995)\citenamefont{Gelfand, Krob,
  Lascoux, Leclerc, Retakh, and Thibon}}]{Gelfand1995}
\bibinfo{author}{\bibfnamefont{I.~M.} \bibnamefont{Gelfand}},
  \bibinfo{author}{\bibfnamefont{D.}~\bibnamefont{Krob}},
  \bibinfo{author}{\bibfnamefont{A.}~\bibnamefont{Lascoux}},
  \bibinfo{author}{\bibfnamefont{B.}~\bibnamefont{Leclerc}},
  \bibinfo{author}{\bibfnamefont{V.~S.} \bibnamefont{Retakh}},
  \bibnamefont{and} \bibinfo{author}{\bibfnamefont{J.-Y.}
  \bibnamefont{Thibon}}, \bibinfo{journal}{Adv. Math.}
  \textbf{\bibinfo{volume}{112}}, \bibinfo{pages}{218} (\bibinfo{year}{1995}).

\bibitem[{\citenamefont{Bateman}(1940)}]{Bateman1940}
\bibinfo{author}{\bibfnamefont{H.}~\bibnamefont{Bateman}},
  \bibinfo{journal}{Proc. N. A. S.} \textbf{\bibinfo{volume}{26}},
  \bibinfo{pages}{491} (\bibinfo{year}{1940}).

\bibitem[{\citenamefont{Cox et~al.}(2007)\citenamefont{Cox, Little, and
  O'Shea}}]{Cox2007}
\bibinfo{author}{\bibfnamefont{D.~A.} \bibnamefont{Cox}},
  \bibinfo{author}{\bibfnamefont{J.}~\bibnamefont{Little}}, \bibnamefont{and}
  \bibinfo{author}{\bibfnamefont{D.}~\bibnamefont{O'Shea}},
  \emph{\bibinfo{title}{Ideals, Varieties, and Algorithms: An Introduction to
  Computational Algebraic Geometry and Commutative Algebra}}
  (\bibinfo{publisher}{Springer-Verlag New York, Inc.},
  \bibinfo{address}{Secaucus, NJ, USA}, \bibinfo{year}{2007}).

\bibitem[{\citenamefont{Buchberger}(2006)}]{Buchberger2006}
\bibinfo{author}{\bibfnamefont{B.}~\bibnamefont{Buchberger}}, Ph.D. thesis
  (\bibinfo{year}{2006}).

\bibitem[{\citenamefont{Buchberger}(1998)}]{Buchberger1998}
\bibinfo{author}{\bibfnamefont{B.}~\bibnamefont{Buchberger}},
  \textbf{\bibinfo{volume}{251}}, \bibinfo{pages}{535} (\bibinfo{year}{1998}).

\bibitem[{\citenamefont{Sturmfels}(2005)}]{Sturmfels2005}
\bibinfo{author}{\bibfnamefont{B.}~\bibnamefont{Sturmfels}},
  \bibinfo{journal}{Notices of the AMS} \textbf{\bibinfo{volume}{52}},
  \bibinfo{pages}{2} (\bibinfo{year}{2005}).

\bibitem[{\citenamefont{Sturmfels}(2002)}]{Sturmfels2002}
\bibinfo{author}{\bibfnamefont{B.}~\bibnamefont{Sturmfels}},
  \emph{\bibinfo{title}{{Solving Systems of Polynomial Equations (Cbms Regional
  Conference Series in Mathematics)}}} (\bibinfo{publisher}{{American
  Mathematical Society}}, \bibinfo{year}{2002}).

\bibitem[{\citenamefont{Dube}(1990)}]{Dube1990}
\bibinfo{author}{\bibfnamefont{T.~W.} \bibnamefont{Dube}},
  \bibinfo{journal}{SIAM J. Comput.} \textbf{\bibinfo{volume}{19}},
  \bibinfo{pages}{750} (\bibinfo{year}{1990}).

\bibitem[{\citenamefont{Bando et~al.}(2013)\citenamefont{Bando, Ichikawa,
  Kondo, and Nakahara}}]{Bando2013}
\bibinfo{author}{\bibfnamefont{M.}~\bibnamefont{Bando}},
  \bibinfo{author}{\bibfnamefont{T.}~\bibnamefont{Ichikawa}},
  \bibinfo{author}{\bibfnamefont{Y.}~\bibnamefont{Kondo}}, \bibnamefont{and}
  \bibinfo{author}{\bibfnamefont{M.}~\bibnamefont{Nakahara}},
  \bibinfo{journal}{J. Phys. Soc. Jpn.} \textbf{\bibinfo{volume}{82}}
  (\bibinfo{year}{2013}).

\bibitem[{\citenamefont{Shaka and Freeman}(1984)}]{Shaka1984}
\bibinfo{author}{\bibfnamefont{A.}~\bibnamefont{Shaka}} \bibnamefont{and}
  \bibinfo{author}{\bibfnamefont{R.}~\bibnamefont{Freeman}},
  \bibinfo{journal}{Journal of Magnetic Resonance (1969)}
  \textbf{\bibinfo{volume}{59}}, \bibinfo{pages}{169 } (\bibinfo{year}{1984}).

\bibitem[{\citenamefont{Cho et~al.}(1986)\citenamefont{Cho, Tycko, Pines, and
  Guckenheimer}}]{Cho1984}
\bibinfo{author}{\bibfnamefont{H.~M.} \bibnamefont{Cho}},
  \bibinfo{author}{\bibfnamefont{R.}~\bibnamefont{Tycko}},
  \bibinfo{author}{\bibfnamefont{A.}~\bibnamefont{Pines}}, \bibnamefont{and}
  \bibinfo{author}{\bibfnamefont{J.}~\bibnamefont{Guckenheimer}},
  \bibinfo{journal}{Phys. Rev. Lett.} \textbf{\bibinfo{volume}{56}},
  \bibinfo{pages}{1905} (\bibinfo{year}{1986}).

\end{thebibliography}

\end{document}